\documentclass[useAMS,usenatbib]{mn2e}
\usepackage[pdftex]{graphicx}
\usepackage[usenames,dvipsnames]{color}
\usepackage{aas_macros}
\usepackage{lastpage}
\usepackage{comment}
\usepackage{subfigure}
\usepackage{hyperref}

\hypersetup{
    pdftitle={A formation history model of galaxy stellar mass growth}, 
    pdfauthor={Simon Mutch},     
    pdfsubject={Article},   
    pdfnewwindow=true      
}

\def\SMF{stellar mass function}
\def\SHMR{stellar--halo mass relation}

\def\FphysMvir{$F_{\rm phys}(M_{\rm vir})$}
\def\FphysVmax{$F_{\rm phys}(V_{\rm max})$}
\def\Vmax{$V_{\rm max}$}
\def\Mvir{$M_{\rm vir}$}
\def\EMvir{$\mathcal{E}_{M_{\rm vir}}$}
\def\EVmax{$\mathcal{E}_{V_{\rm max}}$}
\def\MvirPeak{$M_{\rm peak}$}
\def\VmaxPeak{$V_{\rm peak}$}
\def\sigmaMvir{$\sigma_{M_{\rm vir}}$}
\def\sigmaVmax{$\sigma_{V_{\rm max}}$}
\def\z0{$z{=}0$}

\title[A formation history model of galaxy growth]{The simplest
model of galaxy formation - I. A formation history model of galaxy stellar mass
growth}

\author[Mutch, Croton and Poole]{Simon J. Mutch$^{1,2}$\thanks{E-mail:smutch@unimelb.edu.au}, 
Darren J. Croton$^{2}$
and Gregory B. Poole$^{1}$ \\ 
$^{1}$School of Physics, The University of Melbourne, Parkville, VIC 3010,
Australia\\
$^{2}$Centre for Astrophysics and Supercomputing, 
Swinburne University of Technology, PO Box 218, Hawthorn, VIC 3122, Australia}

\begin{document}


\pagerange{\pageref{firstpage}--\pageref{lastpage}} \pubyear{2013}

\maketitle

\label{firstpage}

\begin{abstract} 
  We introduce a simple model to self-consistently connect the growth of
  galaxies to the formation history of their host dark matter haloes.  Our model
  is defined by two simple functions: the ``baryonic growth function'' which
  controls the rate at which new baryonic material is made available for star
  formation, and the ``physics function'' which controls the efficiency with
  which this material is converted into stars.  Using simple, phenomenologically
  motivated forms for both functions that depend only on a single halo property,
  we demonstrate the model's ability to reproduce the $z{=}0$ red and blue
  stellar mass functions.  Furthermore, by adding redshift as a second input
  variable to the physics function we show that the reproduction of the global
  stellar mass function out to $z{=}3$ is improved.  We conclude by discussing
  the general utility of our new model, highlighting its usefulness for creating
  mock galaxy samples which have a number of key advantages over those generated
  by other techniques.
\end{abstract}


\section{Introduction}
\label{sec:intro}

Theoretical models are an important and commonly used tool for interpreting
and furthering our understanding of observed galaxy populations.  Typically,
these models are used to generate mock galaxy catalogues that can be compared
to equivalent samples drawn from the real Universe.  Knowledge of the models'
construction, combined with their successes and failures in reproducing the
observations, can often allow important inferences to be made about the physics
of galaxy formation and evolution.

There are a number of different methods for generating mock galaxy samples
for comparison with observations. At the most advanced and complex end of the
scale are full hydrodynamic simulations.  These attempt to solve the physics
of galaxy formation from first principles, directly modelling complex baryonic
processes such as cooling and shocks in tandem with the dissipationless growth
of dark matter structure.  Unfortunately, the associated high computational
cost prohibits the resolution of small scale physical processes such as star
formation and black hole feedback in a volume large enough to provide a
cosmologically significant sample of galaxies.  Hence hydrodynamic simulations
often resort to parametrized approximations to deal with these unresolved
``sub-grid'' processes.  In addition they must also deal with the complex and
often poorly understood numerical effects that come with modelling dissipational
physics using finite physical and temporal resolutions.

Semi-analytic galaxy formation models attempt to overcome the computational
costs associated with hydrodynamic simulations by separating the baryonic
physics of galaxy formation from the dark-matter-dominated growth of
structure \citep{White1991,Kauffmann1999}. This is achieved by taking
pre-generated dark matter halo merger trees and post-processing them with
a series of physically motivated parametrizations that attempt to capture
the mean behaviour of the dominant baryonic processes involved in galaxy
formation. The resulting speed means that these models can be used to generate
cosmologically significant samples of galaxies using only modest computing
resources. However, semi-analytic models typically require a number of
free parameters, many of which are often not well constrained by theory or
observation \citep{Neistein2010}. The complicated and degenerate nature of
the different physical prescriptions also means that the effects of these
parameters on the final galaxy population are often highly degenerate and can
be difficult to interpret \citep[e.g.][]{Lu2012}. Additionally, our relatively
poor understanding of high-redshift galaxy formation means that at least some
of the parametrizations used may not be appropriate at these early times
\citep{Henriques2013,Mutch2013}.

For many science questions and applications we are not required (or
able) to include and understand all of the relevant input physics.  In
these cases it is often sufficient to construct simple ``toy'' models
\citep[e.g.][]{Wyithe2007,Dekel2013,Tacchella2013}.  These typically build
an average population of galaxies using simplified approximations, and are
designed to test new ideas and interpretations or to allow the investigation
of particular trends or features found in observational data.  One example is
the ``reservoir'' model of \citet{Bouche2010}.  Here, averaged dark matter
halo growth histories are used to track the typical build up of cold gas
in galaxies.  In their fiducial formalism, the accretion of baryons on to
a galaxy is only allowed to occur when the host halo lies in a fixed mass
range.  However, within this range the accretion is modelled as a simple
fraction of the halo growth rate.  Using a standard Kennicutt--Schmidt law
\citep{Kennicutt1998} for star formation, this simple model is able to
reproduce the observed scaling behaviours of the star-forming main sequence and
Tully--Fisher relations.  Expanding upon this framework, \citet{Krumholz2012}
also introduced a metallicity-dependent star formation efficiency, allowing them
to straightforwardly investigate the associated effects on the star formation
histories of galaxies.

Rather than attempting to generate galaxy populations based on our theories of
the relevant physics, an alternative method of generating mock galaxy samples
is to use purely statistical methods.  Halo occupation distribution (HOD)
models use observed galaxy clustering measurements to constrain the number
of galaxies of a particular type within a dark matter halo of a given mass
\citep{Peacock2000,Zheng2005}.  For the purpose of constructing mock galaxy
catalogues, such a methodology has the advantage that it requires no knowledge
of the how each galaxy forms and is also statistically constrained to produce
the correct result.  However, this limits our ability to learn about the physics
of galaxy evolution, as one has no way to self-consistently connect individual
galaxies at any given redshift to their progenitors or descendants at other
times.

A similar method to HODs for creating purely statistical mock catalogues is
subhalo abundance matching \citep[SHAM; e.g.][]{Conroy2006}. SHAM models are
typically constructed by generating a sample of galaxies of varying masses,
drawn from an observationally determined stellar mass function. Each galaxy is
then assigned to a dark matter halo taken from a halo mass function generated
using an $N$-body simulation. This assignment is made such that the most massive
galaxy is placed in the most massive halo and so on, proceeding to lower and
lower mass galaxies.  Due to possible differences in the formation histories
of haloes of any given mass, an artificial scatter is often added during this
assignment procedure \citep[e.g.][]{Conroy2007,Behroozi2013b,Moster2013}.

By leveraging the use of dark matter merger trees as the source of the halo
samples at each redshift, both HOD and SHAM studies have been able to provide
important constraints on the average build up of stellar mass in the Universe
\citep{Zheng2007,Conroy2009,Moster2013,Behroozi2013b}.  This has allowed
these studies to also draw valuable conclusions about processes such as
the efficiency of star formation as a function of halo mass and the role of
intra-cluster light (ICL) in our accounting of the stellar mass content of
galaxies.  However, both HOD and SHAM models are applied independently at
individual redshifts and do not self-consistently track the growth history of
individual galaxies.  This limits the remit of these models to considering only
the averaged evolution of certain properties over large samples.

Our goal in this work is to present an alternative class of galaxy formation
model which allows us to achieve the ``best of both worlds''; providing both a
self-consistent growth history of each individual galaxy, whilst also minimizing
any assumptions about the physics which drives this growth.  This is achieved
by tying star formation (and hence the growth of stellar mass) to the growth
of the host dark matter halo in $N$-body dark matter merger trees using a simple
but well motivated parametrization that depends only on the properties of
the halo itself.  In this way, we are able to provide a complete formation
history for every galaxy.  The model we present is closely related to that
of \citet{Cattaneo2011}, but with a number of important generalisations that
increase its utility whilst still maintaining a high level of transparency and
simplicity.

This paper is laid out as follows:  In \S\ref{sec:model} we introduce the
framework of our new model.  In particular, \S\ref{sec:baryons} focusses on how
we build up the baryonic content of dark matter haloes, with the practical details
of the model's application outlined in \S\ref{sec:generating_galaxy_pop}.  In
\S\ref{sec:results} we present some basic results, in particular investigating
the model's ability to reproduce the observed galaxy stellar mass function at
multiple redshifts.  In \S\ref{sec:discussion} we discuss our findings as well
as outline the general utility of the model and a number of possible ways in
which it can be extended.  Finally, we present a summary of our conclusions in
\S\ref{sec:conclusions}.

A 1st-year {\it Wilkinson Microwave Anisotropy Probe}
\citep[{\it WMAP}1;][]{Spergel2003} $\Lambda$ cold dark matter ($\Lambda$CDM)
cosmology with $\Omega_{\rm m}{=}0.25$, $\Omega_{\Lambda}{=}0.75$, $\Omega_{\rm
b}{=}0.045$ is utilized throughout this work.  In order to ease comparison with
the observational data sets employed, all results are quoted with a Hubble
constant of $h{=}0.7$ (where $h{\equiv}H_0/100\, \mathrm{km\,s^{-1}Mpc^{-1}}$)
unless otherwise indicated.  Magnitudes are presented using the Vega photometric
system and a standard \citet{Salpeter1955} initial mass function (IMF) is
assumed throughout.

\section{The simplest model of galaxy formation}
\label{sec:model}

\subsection{The growth of structure}

The aim of the model presented in this work is to self-consistently tie the
growth of galaxy stellar mass to that of the host dark matter haloes in as
simple a way as possible.  In order to achieve this we require knowledge of the
properties and associated histories of a large sample of dark matter haloes
spanning the full breadth of cosmic history.  We obtain this in the form of
merger trees constructed from the output of the $N$-body dark matter Millennium
Simulation \citep{Springel2005}.

Using the evolution of over $10^{10}$ particles in a cubic volume with a side
length of 714 Mpc, the Millennium Simulation merger trees track the build up of
dark matter haloes larger than approximately $2.9{\times} 10^{10}\,M_{\sun}$
over 64 temporal snapshots.  These snapshots are logarithmically spaced in
expansion factor between redshifts 127 and 0, with an average separation of
$\sim$200--350 Myr.  Each individual dark matter structure is identified using
a friends-of-friends linking algorithm with further substructures (subhaloes)
identified using the {\small SUBFIND} algorithm of \citet{Springel2001}.  The
simulation employs a concordance $\Lambda{\rm CDM}$ cosmology compatible
with first year {\it WMAP} \citep{Spergel2003} parameters: $(\Omega_{\rm
m},\Omega_{\Lambda},\sigma_8,h_0) = (0.25, 0.75, 0.9, 0.73)$.

\subsection{The baryonic content of dark matter haloes}
\label{sec:baryons}

The maximum star formation rate of a galaxy is regulated by the availability
of baryonic material that can act as fuel.  In our formation history model we
assume that every dark matter halo carries with it the universal fraction of
baryonic material, $f_{\rm b}{=}0.17$ \citep{Spergel2003}. However, some of
these baryons will already be locked up in stars or contained in reservoirs
of material that are unable to participate in star formation.  Therefore we
parametrize the dependence of the amount of newly accreted baryonic material
which is available for star formation on the properties of the host
dark matter halo using a {\it baryonic growth function}, $F_{\rm growth}$.

In practice, only some fraction of this available material will actually
make its way in to the galaxy, with an even smaller amount then successfully
condensing to form stars in a suitably short time interval.  The efficiency
with which this occurs depends on a complex interplay of non-conservative
baryonic processes, both internal and external to the galaxy--halo system.  A
number of important examples include shock heating, feedback from supernova and
active galactic nuclei (AGN), as well as environmental processes such as galaxy
mergers and tidal stripping.  Here we assume that all of these complicated and
intertwined mechanisms can be distilled down into a single, arbitrarily complex
{\it physics function}, $F_{\rm phys}$.

Combining all of this together, we can write down a deceptively simple equation
for the growth rate of stellar mass ($\dot M_{*}$) in the Universe on a per-halo
basis:
\begin{equation}
  \label{eqn:sfr}
  \dot M_{*} = F_{\rm growth} F_{\rm phys}\,.
\end{equation}

In the following sections, we discuss the form we employ for the baryonic growth
and physics functions in turn.

\subsubsection{The baryonic growth function}

In order to explore the simplest form of our formation history model, we begin
by assuming that as a dark matter halo grows, all of the fresh baryonic material
it brings with it is immediately available for star formation.  This corresponds
to a baryonic growth function which is simply given by the rate of growth of the
host dark matter halo:
\begin{equation}
  \label{eqn:bgf}
  F_{\rm growth} = f_{\rm b} \frac{dM_{\rm vir}}{dt}\,.
\end{equation}

\begin{figure}
  \includegraphics[width=\columnwidth]{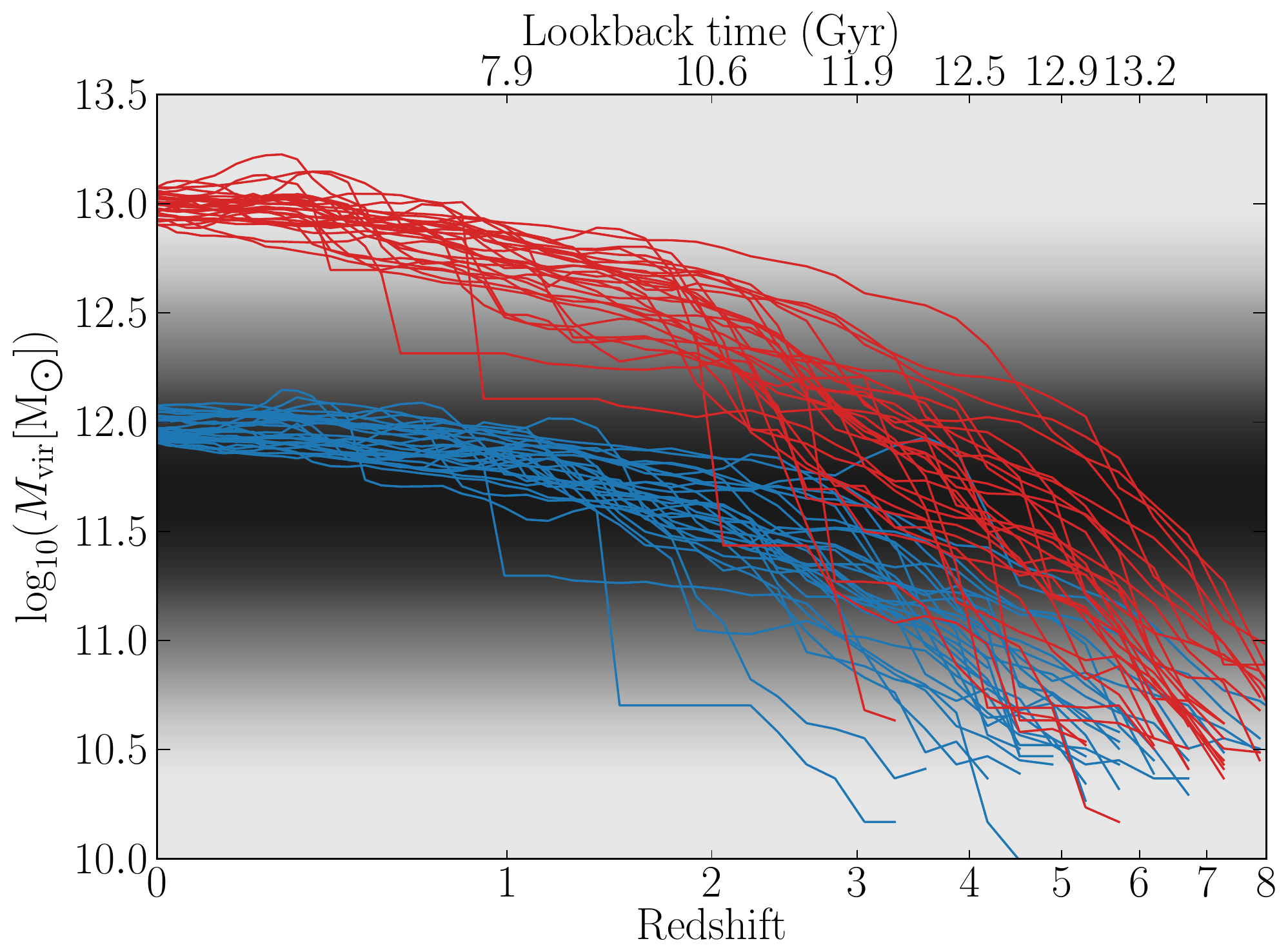}
  \caption{\label{fig:growth_diversity} The large variation in the possible
  growth histories of haloes which all have approximately equal masses by
  redshift zero.  The grey shaded region indicates the amplitude of the physics
  function in Eqn. 3.  The blue and red lines represent 30 randomly selected
  growth histories for haloes with final \Mvir{} values of approximately
  $10^{12}$ and $10^{13}\, M_{\sun}$, respectively.  Variations of 3--4 Gyr
  in the time at which these haloes reach a given mass is common.  Unlike
  statistical techniques for tying galaxy properties to their host haloes, our
  formation history model implicitly includes the full range of different halo
  growth histories and their effects on the predicted galaxy population.}
\end{figure}

In practice haloes of the same $z{=}0$ mass may show a diverse range
of growth histories, all of which are captured by our model.  In
Fig.~\ref{fig:growth_diversity} we demonstrate this by showing the individual
growth histories of a random sample of dark matter haloes selected from the
Millennium Simulation in two narrow mass bins.  From this figure we see that
there can be significant variations in the time at which similar haloes at
redshift zero reach a given mass.  For example, in the upper halo mass sample,
some haloes reach $10^{12}\,{\rm M_{\sun}}$ by $z{=}5$ whilst others do not
reach this value until $z{=}2$.  In addition, some haloes may have complex growth
histories, achieving their maximum mass at $z{>}0$.  This can potentially be
caused by a number of processes such as stripping during dynamical encounters
with other haloes.  Since the baryonic growth function maps the formation history
of each individual dark matter halo to the stellar mass growth of its galaxy,
this diversity in growth histories is fully captured, propagating through to
be reflected in the predicted galaxy populations at all redshifts.  This is an
important attribute of our model that sets it apart from other statistical-based
methods which merely map the properties of galaxies to the instantaneous or
mean properties of haloes, independently of their histories (e.g. HOD and SHAM
models).  These methods typically have to add artificial scatter to approximate
the effects of variations in the halo histories, whereas this variation is a
self-consistent input to our formation history model.

\subsubsection{The physics function}
\label{sec:physics_func}

The physics function describes the efficiency with which baryons are converted
into stars in haloes of a given mass.  The form of this function may be
arbitrarily complex, however, the goal of this work is to find the simplest
model which successfully ties the growth of galaxy stellar mass to the
properties of the host dark matter haloes.  The physics function is not meant
to provide an accurate reproduction of the details of the full input physics,
but rather their combined {\it effects} on the growth of stellar mass in the
Universe.  In this spirit, we begin by assuming that there is only one input
variable: the instantaneous virial mass of the halo, $M_{\rm vir}$.

\begin{figure}
  \includegraphics[width=\columnwidth]{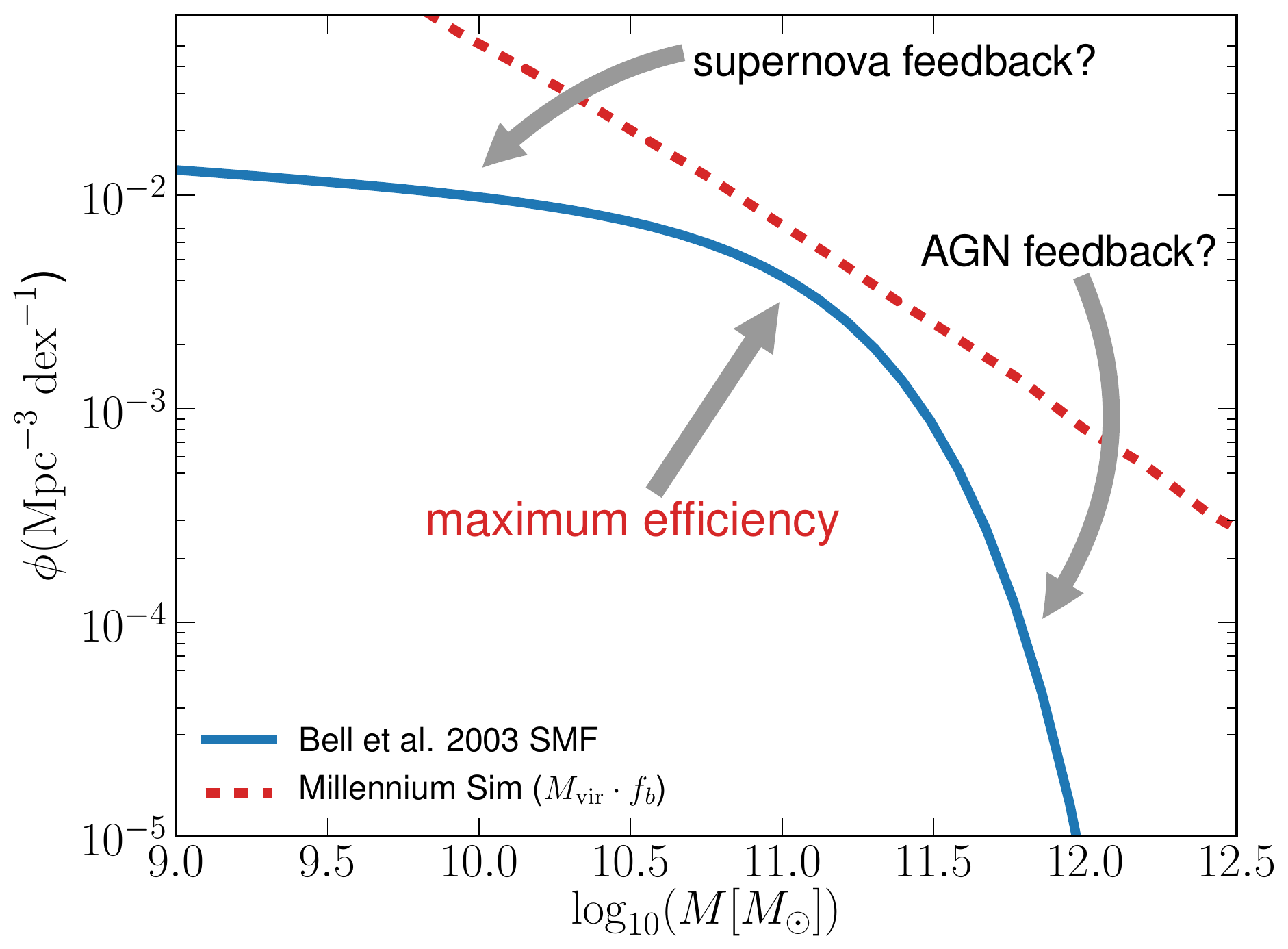}
  \caption{\label{fig:halo_vs_stellar_MFs} A comparison of the observed galactic
    stellar mass function (blue solid line) and the halo mass function
    of the Millennium Simulation (red dashed line).  The halo mass function has
    been multiplied by the universal baryon fraction in order to demonstrate
    the maximum possible stellar mass content as a function of halo mass.
    The closer the stellar mass function is to this line, the more efficient
    star formation is in haloes of the corresponding mass.  If galaxies were
    to form stars with a fixed efficiency at all halo masses then the slope
    of the stellar mass function would be identical to that of the halo mass
    function.  The differing slopes at both high and low masses indicates that star
    formation (as a function of halo mass) is less efficient in these regimes.
    At low masses, this is commonly attributed to efficient gas ejection due to
    supernova feedback, whereas at high masses energy injection from central
    super-massive black holes is thought to be able to effectively reduce the
    efficiency of gas cooling.  However, many other physical processes may also
    contribute in both regimes.}
\end{figure}

Although still not understood in detail, the observed relationship
between dark matter halo mass and galaxy stellar mass is well documented
\citep[e.g.][]{Zheng2007,Yang2012,Wang2013}.  Assuming the favoured $\Lambda
{\rm CDM}$ cosmology, a comparison of the observationally determined galactic
stellar mass function to the theoretically determined halo mass function
indicates that the averaged efficiency of stellar mass growth varies strongly
as a function of halo mass.  In Fig.~\ref{fig:halo_vs_stellar_MFs}, we contrast
a Schechter function fit of the observed redshift zero \SMF{} \citep[solid blue
line;][]{Bell2003} against the dark matter halo mass function of the Millennium
Simulation (red dashed line).  The halo mass function has been multiplied by
$f_{\rm b}$ in order to approximate the total amount of baryons available for
star formation in a halo of any given mass.

The increased discrepancy between the stellar mass function and halo mass
functions at both low and high masses indicates that the efficiency of star
formation is reduced in these regimes.  It is commonly held that at low masses
the shallow gravitational potential provided by the dark matter haloes allows
supernova feedback to efficiently eject gas and dust from the galaxy.  This
reduces the availability of this material to fuel further star formation
episodes, hence temporarily stalling in situ stellar mass growth.  Other
processes such as the photoionization heating of the intergalactic medium
may also play an important role in reducing the efficiency of star formation
in this low-mass regime \citep[][and references therein]{Benson2002}.  At
high halo masses, it is thought that inefficient cooling coupled with strong
central black hole feedback also leads to a quenching of star formation
\citep[e.g.][]{Croton2006}.  Therefore, it is only between these two extremes,
around the knee of the galactic stellar mass function, that stellar mass growth
reaches its highest average efficiency.

We begin by parametrizing the physics function as a simple log-normal
distribution centred around a halo virial mass $M_{\rm peak}$, and with a
standard deviation $\sigma_{M_{\rm vir}}$:
\begin{equation}
  \label{eqn:physicsfunc_mvir}
  F_{\rm phys}(M_{\rm vir}/M_{\sun}) =
      \mathcal{E}_{M_{\rm vir}} \exp\left(-\left(\frac{\Delta M_{\rm vir}}{\sigma_{M_{\rm
      vir}}}\right)^2\right),
\end{equation}
where $\Delta M_{\rm vir} {=} \log_{10}(M_{\rm
vir}/M_{\sun}){-}\log_{10}(M_{\rm peak}/M_{\sun})$ and the parameter \EMvir{}
represents the maximum possible efficiency for converting in-falling baryonic
material into stellar mass, achieved when $M_{\rm vir}{=}M_{\rm peak}$.  Such
a distribution has been found by SHAM studies to provide a good match to the
derived star formation rates as a function of halo mass for $z{\la}2$
\citep{Conroy2009,Bethermin2012}.

\begin{figure*}
  \begin{minipage}{\textwidth}
    \begin{center}
      \includegraphics[width=0.8\columnwidth]{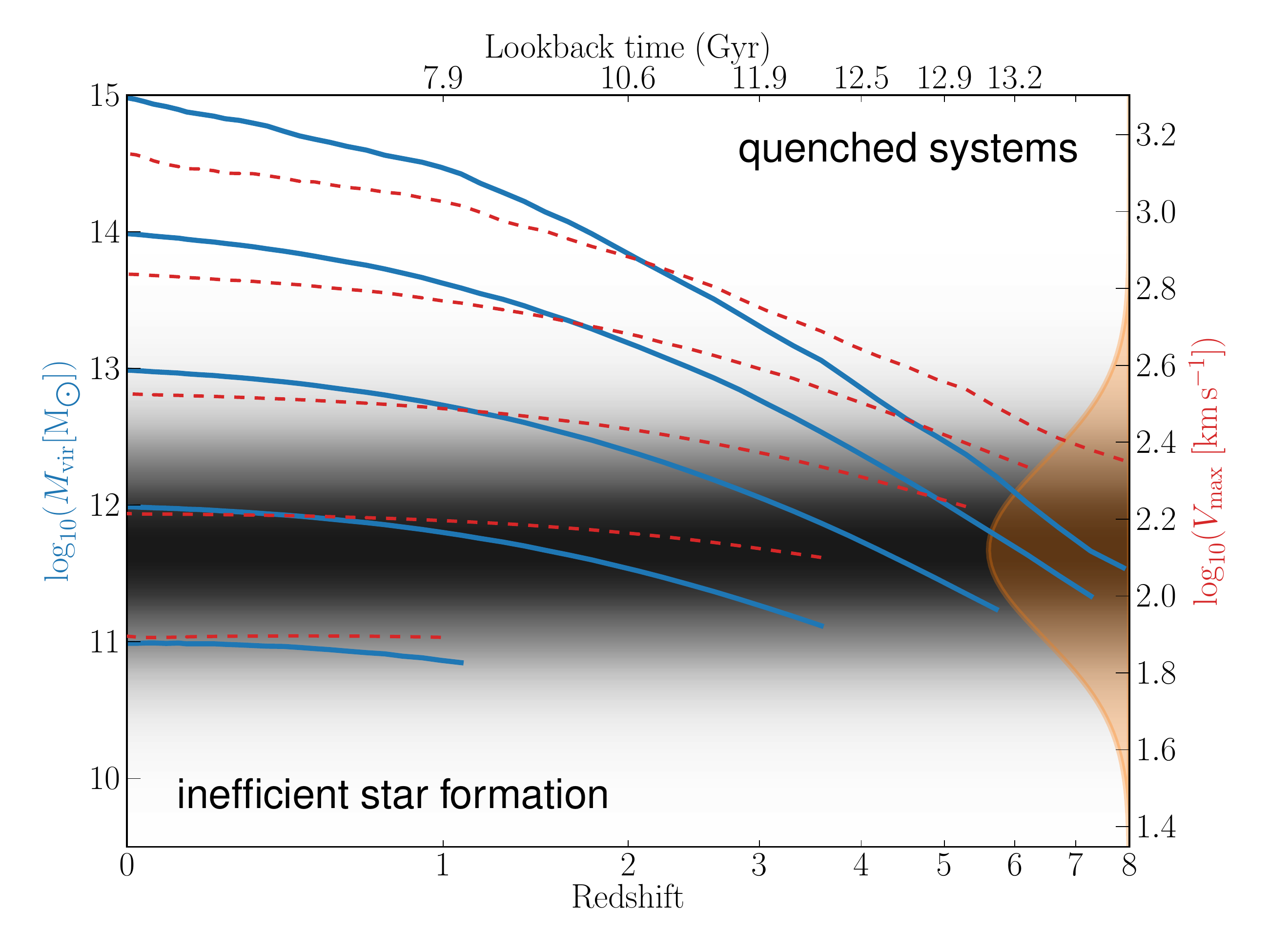}
      \caption{\label{fig:cartoon} The mean virial mass (\Mvir{}) growth
 histories for five samples of dark matter haloes with varying final masses
 (blue solid lines).  Each sample is only plotted out to a redshift limit
 determined by where 80\% of the haloes contain more than 40 particles.  The
 grey shaded region indicates the amplitude of the physics function in
 Eqn.~\ref{eqn:physicsfunc_mvir}.  This is also illustrated by the log-normal
 curve on the right-hand side (orange shaded region). Galaxies with halo masses
 within the peaked region form stars efficiently; outside this mass range, the
 amount of star formation is negligible.  All galaxies of sufficient mass at
 \z0{} will cross the efficient star formation mass range at some point in their
 history, with this period typically coming earlier for more massive haloes. Also
 shown is the mean maximum circular velocity (\Vmax{}) evolution for haloes of
 varying final velocity (red dashed lines).  These samples have been chosen
 to have \z0{} maximum circular velocity values similar to the mean values of
 the five mass-selected samples.  There are clear differences in the evolution
 of $V_{\rm max}$ and $M_{\rm vir}$ which will in turn result in differences
 between the produced galaxy populations.}
    \end{center}
  \end{minipage}
\end{figure*}

This simple form of the physics function provides a number of desirable
properties.  In Fig.~\ref{fig:cartoon}, we present the average growth histories
of five samples of dark matter haloes chosen from the Millennium Simulation
merger trees by their final redshift zero masses (solid blue lines).  For
clarity, we only plot these histories out to redshifts where more than 80\%
of the haloes in each sample have masses which are twice the resolution limit
of the input merger trees.  The grey shaded region indicates the amplitude
of the physics function defined by Eqn~\ref{eqn:physicsfunc_mvir} when using
our fiducial parameter values (see \S\ref{sec:z0_results} for details).  As
the haloes grow, they pass through the region of efficient star formation at
different times depending on their final masses.  Galaxies hosted by the most
massive \z0{} haloes form the majority of their in situ stellar mass at earlier
times whereas those in the lowest mass haloes are still to reach the peak of
their growth.  In addition, lower mass haloes tend to spend a longer time in the
efficient star forming regime compared to their high-mass counterparts.  These
trends qualitatively agree with the observed phenomenon of galaxy downsizing
\citep[e.g.][]{Cowie1996,Cattaneo2008}.

Subhalo abundance matching studies have suggested that $V_{\rm max}$ may be
more tightly coupled to the stellar mass growth of galaxies than $M_{\rm vir}$
\citep[e.g.][]{Reddick2013}.  This makes intuitive sense as $V_{\rm max}$ is
directly related to the gravitational potential of the inner regions of the host
halo, where galaxy formation occurs.  Therefore, in addition to virial mass we also
consider the case of a physics function where the dependent variable is the
instantaneous maximum circular velocity of the host halo, \Vmax{}:
\begin{equation}
  \label{eqn:physicsfunc_vmax}
  F_{\rm phys}(V_{\rm max}/{\rm (km\,s^{-1})}) =
      \mathcal{E}_{V_{\rm max}} \exp\left(-\left(\frac{\Delta V_{\rm max}}{\sigma_{V_{\rm
      max}}}\right)^2\right),
\end{equation}
where $\Delta V_{\rm max} {=} \log_{10}(V_{\rm max}/{\rm
(km\,s^{-1})}){-}\log_{10}(V_{\rm peak}/{\rm (km\,s^{-1})})$.  To avoid
confusion, from now on we will refer to the formation history model constructed
using this physics function as the ``static \Vmax{} model''.  Similarly, we will
refer to the case of \FphysMvir{} as the ``static \Mvir{} model''.

In Fig.~\ref{fig:cartoon} we show the average $V_{\rm max}$ growth histories for
a number of different \z0{} selected samples.  The $y$-axis has been scaled such
that the grey band also correctly depicts the changing amplitude of the \Vmax{}
physics function as well as its \Mvir{} counterpart.  Additionally, each of the
\Vmax{} samples in Fig.~\ref{fig:cartoon} (red dashed lines) is chosen to have
mean \z0{} values close to that of the five $M_{\rm vir}$ samples (blue lines).
However, there are clear differences between the growth histories of these two
halo properties.  In particular, the evolution of \Vmax{} is slightly flatter,
resulting in haloes transitioning out of the efficient star forming region at an
earlier time than the equivalent $M_{\rm vir}$ sample.  Such differences will
have important consequences for the time evolution of the galaxy populations
generated by each of the two physics functions and we highlight some of these in
\S\ref{sec:results}.

By combining the baryonic growth function with a physics function of the forms
presented here, our resulting model may be thought of as a simplified and
extended version of that presented by \citet{Bouche2010}.  Unlike their model,
the scaling of gas accretion efficiency with halo mass, and the dependence
of star formation on previously accreted material, is implicitly contained
within our physics function.  Most importantly though, \citet{Bouche2010} use
statistically generated halo growth histories instead of simulated merger trees.
Hence their model contains no information about the scatter due to variations
in halo formation histories.  Furthermore, since their growth histories do not
include satellites, there is no self-consistent stellar mass growth due to
mergers.

The model of \citet{Cattaneo2011} also uses simulated merger trees as input
and thus shares many of the same advantages as our formation history model.
However, their model ties the properties of galaxies to the instantaneous
properties of their host haloes alone.  In contrast, we use the full information
of the mass accretion history to describe the availability of baryonic material
for star formation.  Also, we make no attempt to motivate the precise form of
our model in terms of combinations of particular physical processes and their
scalings with halo properties, as \citet{Cattaneo2011} do.  This allows our
model to remain maximally general and flexible.

\subsection{Generating the galaxy population}
\label{sec:generating_galaxy_pop}

Armed with the forms of our baryonic growth function (Eqn.~\ref{eqn:bgf}) and
physics function (Eqn.~\ref{eqn:physicsfunc_mvir}), we now discuss the practical
implementation of the formation history model to generate a galaxy population
from the input dark matter merger trees.

For each halo in the tree, the change in dark matter halo mass, coupled with
the time between each merger tree snapshot, provides us with the value of ${\rm
d}M_{\rm vir}/{\rm d}t$.  This change in mass naturally includes growth due to
both smooth accretion {\it and} merger events.  Combined with the instantaneous
value of \Mvir{} or \Vmax{} we can calculate a star formation rate for the
occupying galaxy following Eqn.~\ref{eqn:sfr}.

Some fraction of the mass formed by each new star formation episode will be
contained within massive stars.  The lives of these stars will be relatively
short and therefore they will not contribute to the measured total stellar
mass content of the galaxy.  In order to model this effect we invoke the
``instantaneous recycling'' approximation \citep{Cole2000}, whereby some fraction
of the mass of newly formed stars is assumed to be instantly returned to the
galaxy interstellar medium (ISM).  Based on a \citet{Salpeter1955} IMF we take
this fraction to be 30\%, however, we note that changes to this value can be
trivially taken into account by appropriately scaling the value of $\mathcal{E}$
in the physics function.

Although well motivated and conceptually simple, our use of ${\rm d}M_{\rm
vir}/{\rm d}t$ in the baryonic growth function (Eqn.~\ref{eqn:bgf}) does
introduce some practical considerations.  For example, the change in halo
mass from snapshot-to-snapshot in the input dark matter merger trees can
be stochastic in nature, especially for the case of low-mass or diffuse
haloes identified in regions of high density.  Also, when satellite galaxies
fall into larger systems their haloes are tidally stripped, leading to a
negative change in halo mass and thus a reduction in stellar mass according
to Eqn.~\ref{eqn:sfr}.  In the real Universe, we expect that the galaxy is
located deep within the potential well of its host halo and is therefore
largely protected from the earliest stripping effects suffered by the dark
matter \citep{Penarrubia2010}.  We must therefore decide when, if at all,
to allow stellar mass loss when using this formalism.  For simplicity, we
address this by setting the star formation rate of satellite galaxies to
be zero at all times; in other words fixing their stellar mass upon in
fall.  This is unlikely to be true in the real Universe across all mass
and environment scales \citep[][]{Weinmann2006}, however, the assumption
of little or no star formation in satellite galaxies is a reasonable
approximation and is relatively common in analytic galaxy formation models
\citep[e.g.][]{Kauffmann1999,Cole2000,Bower2006,Croton2006}.  It is also in
keeping with our goal of finding the simplest possible model.

The form of the baryonic growth function presented in Eqn.~\ref{eqn:bgf} above
is only one of a number of possibilities.  As an example, one could use the
instantaneous halo mass divided by its dynamical time, $M_{\rm vir}/t_{\rm dyn}$.
This quantity grows more smoothly over the lifetime of a halo and is never
negative.  Additionally, one may speculate that this is a better representation of
the link between stellar and halo mass build up.  However, for simplicity, we do
not investigate alternative forms of the baryonic growth function, but leave
this to future work.

Satellite galaxies are explicitly tracked in the input merger trees until their
host subhaloes can no longer be identified or fall below the imposed resolution
limit of 20 particles.  At this point, their position is approximated by the
location of the most bound particle at the last snapshot the halo was
identified.  We then follow \citet{Croton2006} in assuming that the associated
satellite galaxy will merge with the central galaxy of the parent halo/subhalo
after a time-scale motivated by dynamical friction arguments \citep{Binney2008}:
\begin{equation}
  t_{\rm merge} = \frac{1.17}{G} 
  \frac{V_{\rm vir}r^2_{\rm sat}}{m_{\rm sat}\ln(1+M_{\rm vir}/m_{\rm sat})}\ ,
\end{equation}
where $V_{\rm vir}$ and $M_{\rm vir}$ are the virial velocity and mass of the
parent dark matter halo in $\rm km\,s^{-1}$ and $\rm M_{\sun}$ respectively,
$r_{\rm sat}$ is the current radius of the satellite halo in $\rm kpc$, and
$m_{\rm sat}$ is the mass of the satellite in $\rm M_{\sun}$.  In these
units, the gravitational constant, $G$, is given by $\rm 4.40\times 10^{-9}\:
kpc{^2}\,km\,s^{-1}\,M_{\sun}^{-1}\,Myr^{-1}$.

The final stellar mass of a merger remnant is given by the sum of the stellar
masses of the two merging progenitor galaxies.  This is a key feature of
the model and allows the growth histories of the progenitors of each galaxy
to affect the final stellar populations of their descendants.  Our input
dark matter merger trees are constructed such that the mass of central
friends-of-friends haloes implicitly includes the mass of all of its associated
subhaloes.  Therefore, as an in-falling satellite halo crosses the virial radius
of its parent, the satellite mass is instantaneously added to that that of the
parent, thus contributing to its $dM_{\rm vir}/dt$ value and the amount of star
formation in the central galaxy.  In reality this burst of newly formed stars
could be produced by a number of different mechanisms.  These include star
formation in the central galaxy fuelled by external smooth accretion or material
stripped from the infalling satellite (e.g. hot halo gas), star formation in
the satellite galaxy as it uses up its remaining cold gas reserves during
infall and the merger-driven starburst which may occur when the central or
satellite galaxies eventually do collide and merge.  However, our simple model
makes no assumptions about what contribution each of these mechanisms makes to
the total amount of stars produced during a merger event.  

Combined with our simple baryonic growth function that assumes {\it all} of the
incoming baryonic material is available for star formation (irrespective of
whether or not it is already locked up in stars), our model implicitly includes
merger-driven starbursts with an increased efficiency.  However, since we do not
explicitly account for the amount of incoming baryons which are already locked
up in stars when a satellite halo in-falls, there is the possibility that the
resulting merger-driven star burst produces a system with a baryon fraction in
excess of the universal value.  For our best-fitting models below, we have found
that this situation occurs in less than 0.25\% of all friends-of-friends haloes
at any single snapshot.  The situation is most prevalent in haloes with $M_{\rm
vir}{\approx} 11.5\:M_{\sun}$, although even there, less than 1\% have baryon
fractions above the cosmic value.
 
Knowledge of the star formation rates of each galaxy and its progenitors at
every time step in the simulation allows us to also calculate luminosities.  For
this purpose we use the simple stellar population models of \citet{Bruzual2003}
and assume a \citet{Salpeter1955} IMF.  In the real Universe supernova ejecta
enriches the intra galactic medium, altering the chemical composition of the
next generation of stars and the spectrum of the light they emit.  As we do
not track the amount of gas or metals in our simple model, we assume all stars
are of 1/3 solar metallicity.  This is a common assumption when no metallicity
information is available.  Finally, a simple ``plane-parallel slab'' dust model
\citep{Kauffmann1999} is applied to the luminosity of each galaxy in order to
provide approximate dust extincted magnitudes. These magnitudes are used below
to augment our analysis by allowing us to calculate the $B{-}V$ colour for each
galaxy at \z0{}. However, our main focus will remain on stellar masses as these
are a direct model prediction.

\section{Results}
\label{sec:results}

Having outlined the methodology and implementation of our simple formation
history model, we now present some basic results which showcase its ability
to recreate observed distributions of galaxy properties.  We begin by
considering redshift zero alone, before moving on to investigate the results
at higher redshifts.  Throughout, we contrast the variations between the
predicted galaxy populations when using \Mvir{} or \Vmax{} as the dependant
variable of the physics function (Eqns.~\ref{eqn:physicsfunc_mvir} \&
\ref{eqn:physicsfunc_vmax}).

\subsection{Redshift zero}
\label{sec:z0_results}

\begin{table*}
  \begin{minipage}{\textwidth}
    \begin{center}
  \caption{\label{tab:params} The fiducial parameter values of the physics
  function when using either \Mvir{} or \Vmax{} as the dependant variable.
  Values are presented for both the non-evolving (see \S\ref{sec:z0_results})
  and evolving (see \S\ref{sec:evo_results}) form.  Also shown are
  the ranges for the flat priors used during the MCMC calibration.  The parameters
  $\log_{10}(M_{\rm peak}/\rm{M_{\sun}})$ and $\log_{10}(V_{\rm peak}/({\rm
  km\,s^{-1})})$ indicate the haloes which possess the peak star formation
  efficiency (at \z0{} in the evolving case).  The $\sigma$ and $\mathcal{E}$
  parameters represent the width and height of the Gaussian physics function,
  respectively.  For the evolving models, $\alpha$, $\beta$ and $\gamma$
  indicate the rate of power-law evolution of the peak location, width and
  height of the physics function, respectively.  The non-evolving model
  parameters were chosen to provide the best reproduction of the observed
  $z{=}0$ colour-split stellar mass function of \citet{Bell2003}.  In the
  evolving case, the values were chosen to additionally reproduce the evolution
  of the peak stellar--halo mass relation of \citet{Moster2013}.}
  \center
  \begin{tabular}{|c|c|c|c|c|c|c|}
    \hline
    $\mathbf{M_{\rm vir}}$ {\bf model} &
    $\log_{10}(M_{\rm peak}/\rm{M_{\sun}})$ & $\sigma_{M_{\rm vir}}$ &
    $\mathcal{E}_{M_{\rm vir}}$ & $\alpha_{M_{\rm vir}}$ & $\beta_{M_{\rm
    vir}}$ & $\gamma_{M_{\rm vir}}$\\
    \hline
    Prior ranges & [$11.15, 12.65$] & [$0.2, 1.5$] & [$0.1, 1.5$] & [$-3.0, 3.0$]
    & [$-3.0, 3.0$] & [$-3.0, 3.0$] \\
    \hline
    {\bf Static} {\it(\S\ref{sec:z0_mvir})} &
    $11.7$ & $0.65$ & $0.56$ & -- & -- & --\\ 
    {\bf Evolving} {\it(\S\ref{sec:evo_results})} &
    $11.6$ & $0.56$ & $0.90$ & $0.03$ & $0.25$ & $-0.74$ \\ 
    \hline\hline
    $\mathbf{V_{\rm max}}$ {\bf model} &
    $\log_{10}(V_{\rm peak}/({\rm km\,s^{-1})})$ & $\sigma_{V_{\rm max}}$ & 
    $\mathcal{E}_{V_{\rm max}}$ 
    & $\alpha_{V_{\rm max}}$ &
    $\beta_{V_{\rm max}}$ & $\gamma_{V_{\rm max}}$\\
    \hline
    Prior ranges & [$1.5, 3.5$] & [$0.1, 1.5$] & [$0.1, 1.5$] & [$-3.0, 3.0$]
    & [$-3.0, 3.0$] & [$-3.0, 3.0$] \\
    \hline
    {\bf Static} {\it(\S\ref{sec:z0_vmax})} &
    $2.2$ & $0.18$ & $0.53$ & -- & -- & -- \\ 
    {\bf Evolving} {\it(\S\ref{sec:evo_results})} &
    $2.1$ & $0.17$ & $1.12$ & $0.10$ & $0.33$ & $-0.98$ \\ 
    \hline
  \end{tabular}
  \end{center}
  \end{minipage}
\end{table*}

In order to determine the ``best'' parameter values for the \Mvir{} model, we
calibrate them against Schechter function fits of the observed red and blue
galaxy stellar mass functions of \citet{Bell2003}.  This calibration was done
using Markov chain Monte Carlo (MCMC) parameter estimation techniques \citep[for
details of our implementation see][]{Mutch2013}.  The observed mass functions
are constructed from a $g$-band limited sample taken from a combination of Sloan
Digital Sky Survey (SDSS) early release \citep{Stoughton2002} and Two Micron
All Sky Survey \citep[2MASS;][]{Jarrett2000} data, with a magnitude-dependent
colour cut used to divide the red and blue galaxy populations.  To similarly
split the model galaxies into red and blue samples we employ a more basic
mass/magnitude independent colour cut of $B{-}V=0.8$.  This is equivalent to the
colour division found by the 2dF Galaxy Redshift Survey \citep{Cole2005}.

For all of the results presented in this work we use a minimum of 130\,000
model calls in the integration phase of our Monte Carlo chains, where
the precise number used varies in proportion to the number of free model
parameters.  Due to computational limitations we are unable to utilize
the full Millennium Simulation volume and instead restrict ourselves
to a random sampling of 1/128 of the total simulation merger trees.
This is equivalent to a comoving volume of approximately $9.8\times
10^{6}\mathrm{h}^{-3}\,\mathrm{Mpc^3}$ (i.e. a box with a side length of
approximately $100 \mathrm{h}^{-1}\,\mathrm{Mpc}$).  We use the same random
merger tree set throughout.  Flat priors were used for each parameter with
ranges as presented in Table~\ref{tab:params}.  To ensure that all of our chains
are fully converged we employ the Rubin--Gelman statistic \citep{Gelman1992} as
well as visually inspect the chain traces.

It is important to note that, since our model relies on both the instantaneous
host dark matter halo mass and its growth rate, the precise best-fitting parameter
values may vary depending on the time-step spacing between the snapshots of the
input simulation. For this reason one should be cautious not to over-interpret
the exact parameter values of our simple model as they can be sensitive to
the details of the implementation\footnote{Conversely, our method allows one
to easily investigate the ramifications of varying simulation and merger tree
properties, providing a direct check of such previously hidden differences.}.

Using the posterior probability distributions of the MCMC fitting procedure
allows us place 68 and 95\% confidence limits on all of our model results, both
constrained and predicted.  In this work, confidence intervals are calculated
from a large sample of model runs (60--200) that have parameter combinations
randomly sampled from the relevant posterior distributions.

An important consideration when statistically calibrating any model against
observational data is the use of realistic observational uncertainties
\citep{Mutch2013}.  As discussed by \citet{Bell2003}, there is likely
significant systematic uncertainties associated with their stellar mass
function estimation which are not formally included in the relevant Schechter
function parameter values.  To overcome this we utilize the uncertainties of
\citet{Baldry2008} which are calculated by comparing the global mass functions
that result from five independent stellar mass determinations of a single
galaxy sample.  We then partition this global uncertainty between red and blue
galaxies.  This is done such that the fractional contribution to the uncertainty
due to red (blue) galaxies is equal to the fraction of red (blue) galaxies in
each stellar mass bin.

\begin{figure}
  \includegraphics[width=\columnwidth]{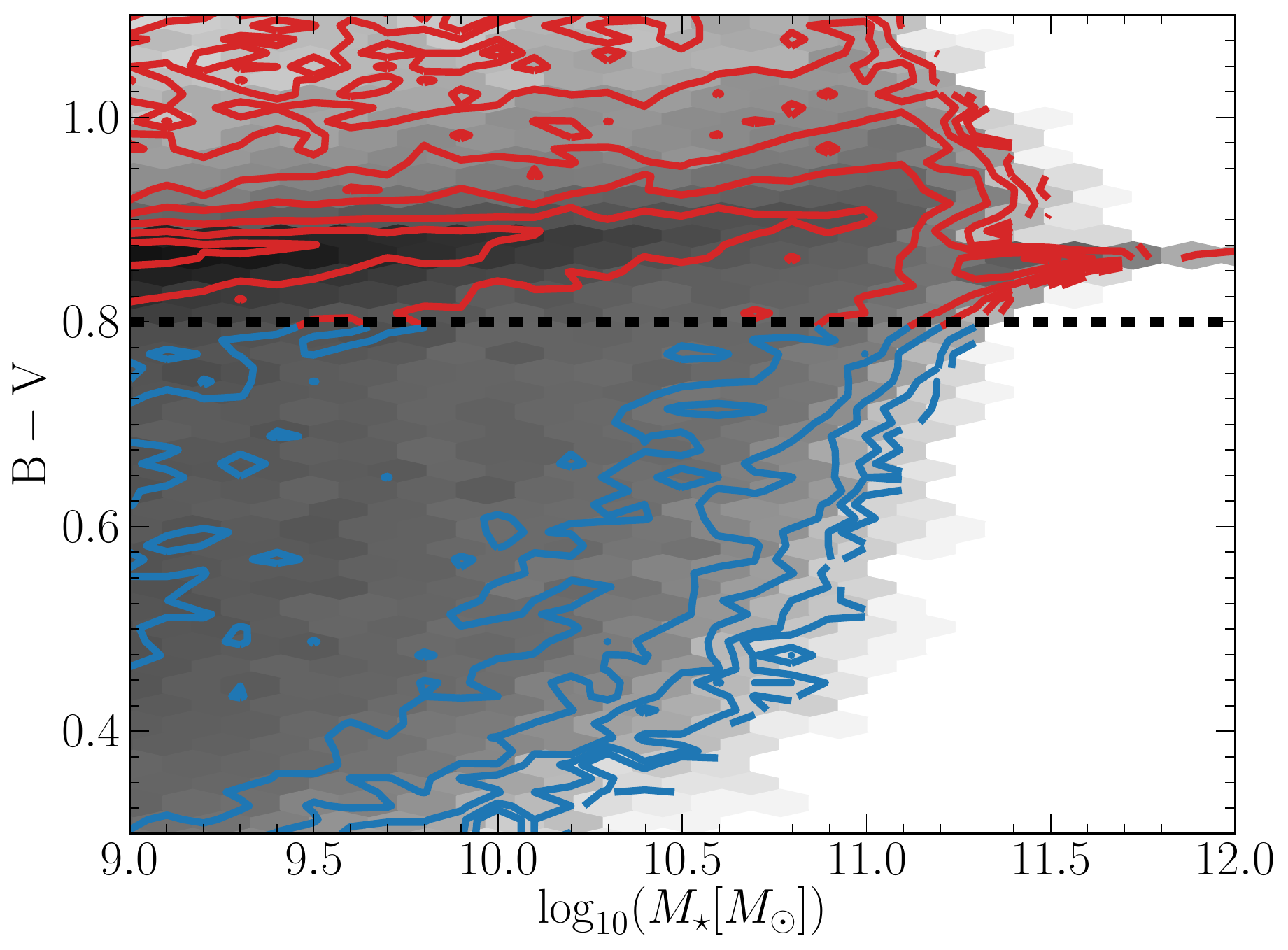}
  \caption{\label{fig:mvir_csmd} The colour--stellar mass diagram of the
  static \Mvir{} model.  The black dashed line at $B{-}V$=0.8 \citep{Cole2005}
  indicates the value used to divide the model galaxies into red and blue
  samples.  There is a clear ridge of over-density extending across all stellar
  masses at $B{-}V{\approx}0.87$ representing the red sequence.}
\end{figure}

\begin{figure*}
  \begin{minipage}{\textwidth}
    \begin{center}
    \subfigure{\includegraphics[width=0.475\textwidth]{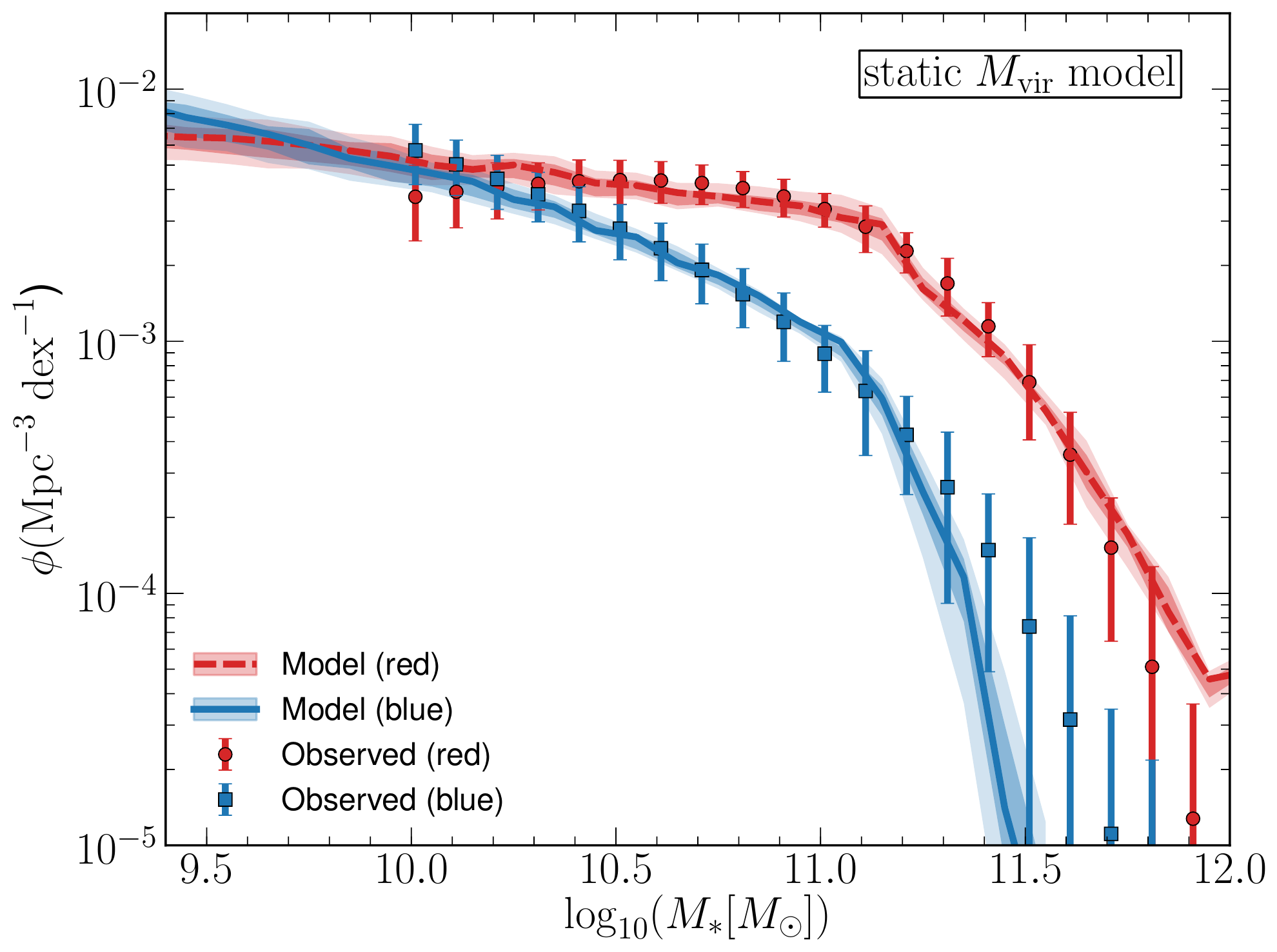}}\quad
    \subfigure{\includegraphics[width=0.475\textwidth]{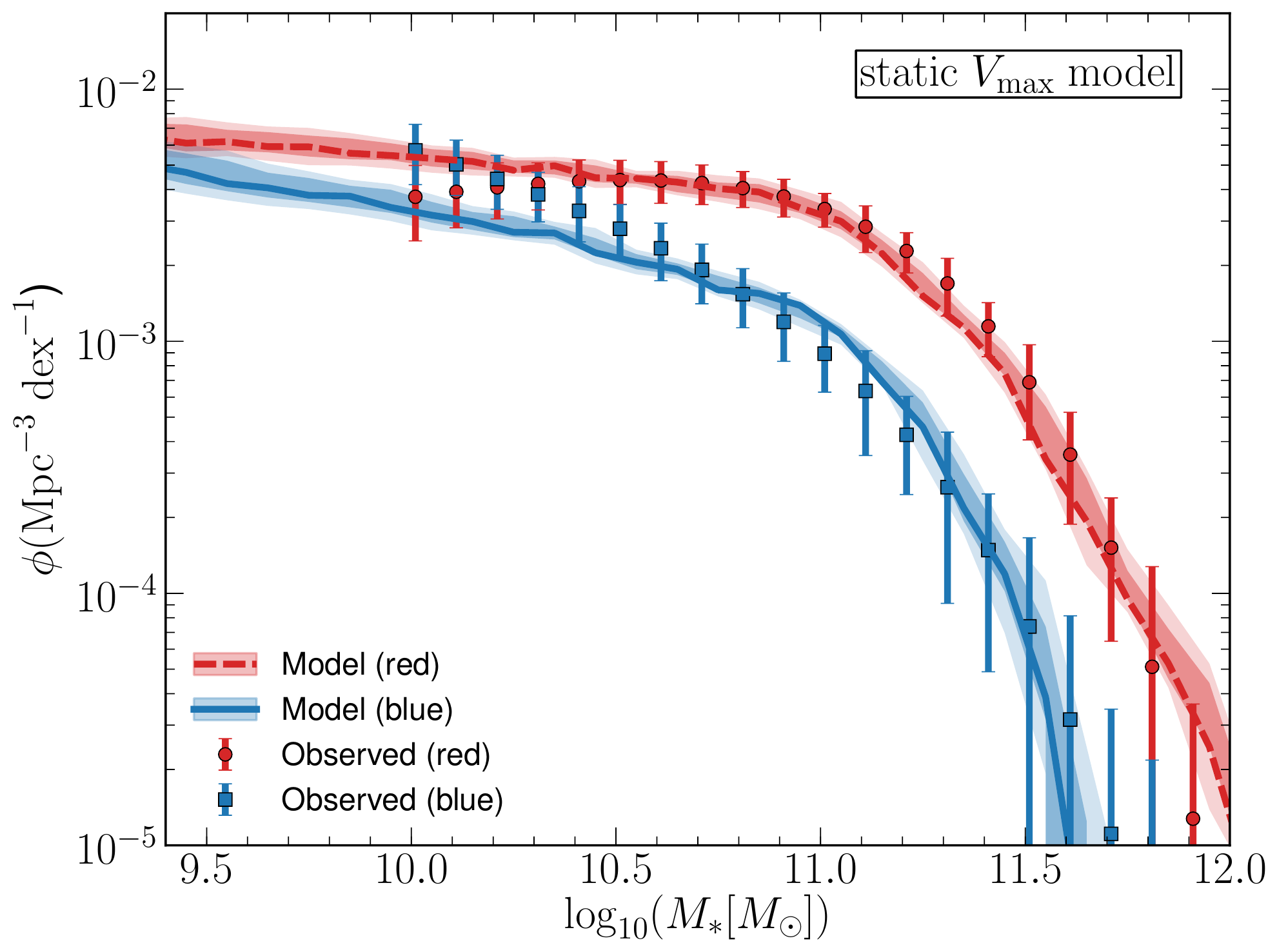}}
    \caption{\label{fig:z0_smf} The red (dashed lines) and blue (solid lines)
    $z{=}0$ \SMF{}s produced by the static formation history model when
    using \Mvir{} (left-hand panel) and \Vmax{} (right-hand panel) as the
    input variable to the physics function (Eqns.~\ref{eqn:physicsfunc_mvir}
    \& \ref{eqn:physicsfunc_vmax}).  Galaxy colour is classified using a
    mass-independent colour cut of $B{-}V{=}0.8$.  The free model parameters
    have been calibrated to provide the best possible reproduction of the
    observed \SMF{} of \citet{Bell2003} (error bars) and are presented in
    Table~\ref{tab:params}.  Shaded regions indicate the 68 (dark) and 95
    (light) \% confidence intervals of our MCMC fit.  The fact that such
    an agreement can be achieved is an important success given the simplicity of
    the formation history model.}
    \end{center}
  \end{minipage}
\end{figure*}

\begin{figure*}
  \begin{minipage}{\textwidth}
    \begin{center}
    \subfigure{\includegraphics[width=0.475\textwidth]{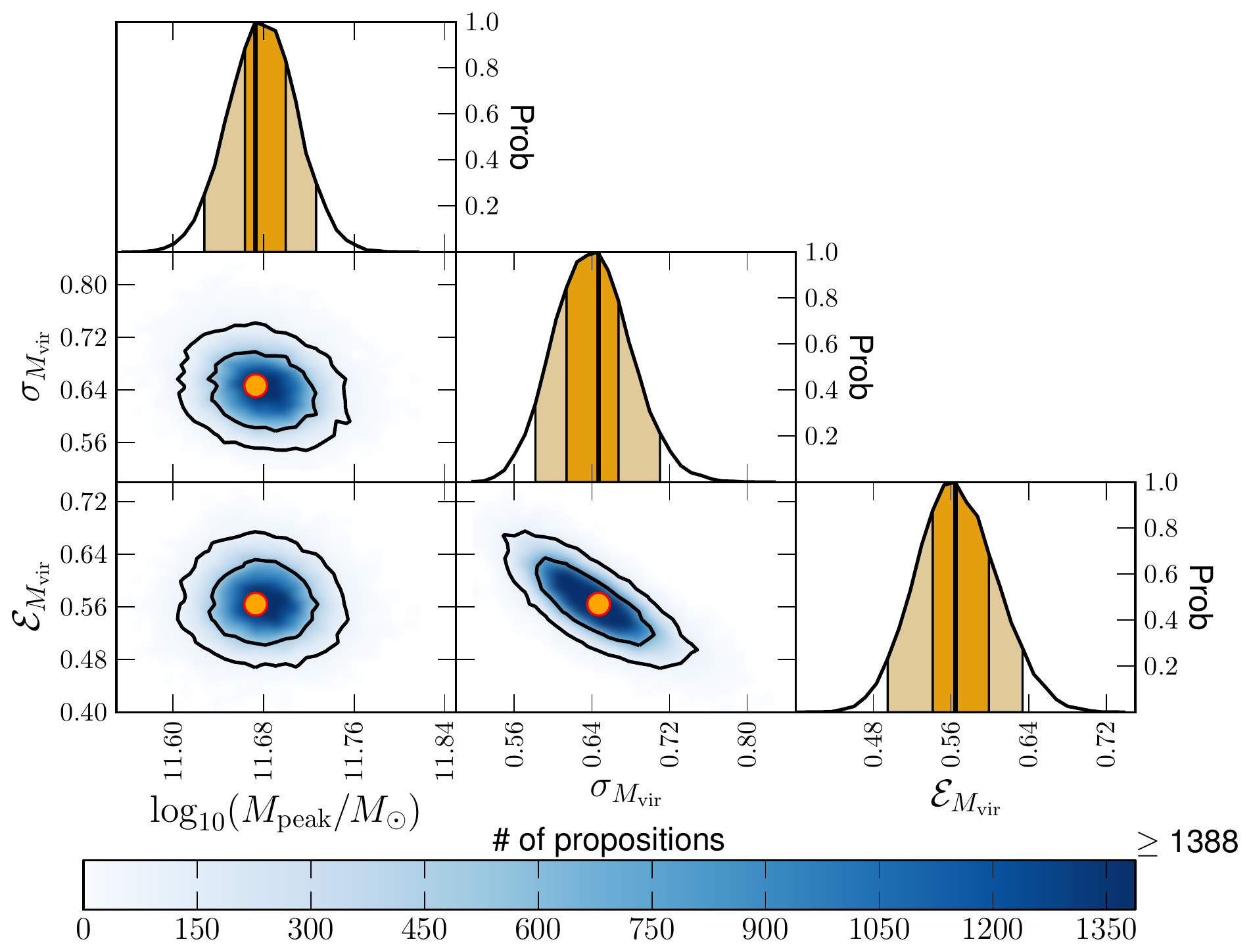}}\quad
    \subfigure{\includegraphics[width=0.475\textwidth]{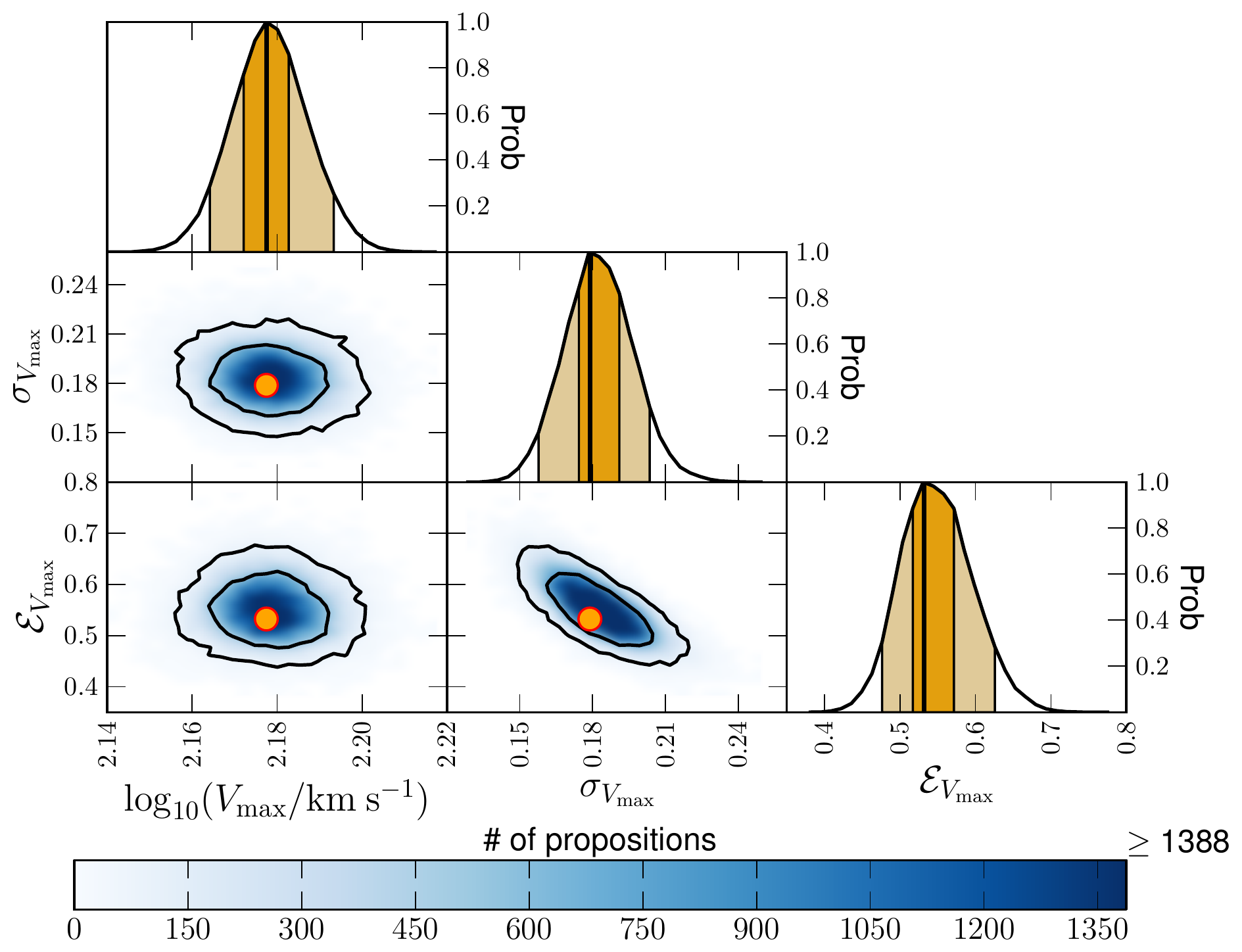}}
    \caption{\label{fig:z0_smf_probs} Marginalized posterior probability
  distributions for the \Mvir{} (left) and \Vmax{} (right) static model
  parameters when calibrating against the red and blue stellar mass functions
  at $z{=}0$ (Fig.~\ref{fig:z0_smf}).  All panels have been zoomed in to the
  high-probability regions.  Contours on the 2D (blue) panels indicate the
  68 and 95\% confidence regions.  Yellow dots mark the marginalized
  most probable parameter values.  The diagonal panels show the marginalized
  1D distributions with the 68 and 95\% confidence intervals (dark and light
  shaded regions, respectively).  The approximately Gaussian shape of these
  1D distributions indicates the well behaved nature of the model.  The only
  parameter degeneracies are between the normalization (\EMvir{}/\EVmax{}) and
  width (\sigmaMvir{}/\sigmaVmax{}) of the physics functions.}
    \end{center}
  \end{minipage}
\end{figure*}

\subsubsection{The \Mvir{} model}
\label{sec:z0_mvir}

The best-fitting parameters for the static \Mvir{} model are presented in
Table~\ref{tab:params} (see \S\ref{sec:evo_results} for the ``evolving''
model).  The preferred \MvirPeak{} value of $10^{11.7}\,\rm{M_{\sun}}$
implies that galaxies in haloes slightly less massive than that of the Milky
Way \citep[$M_{\rm vir}\approx 10^{12}\, \rm{M_{\sun}}$;][]{Xue2008} are on
average the most efficient star formers.  In these haloes, 56\% of all freshly
accreted baryonic material is converted into stars, as indicated by the value
of \EMvir{}.  We also note that both the position of the peak of the physics
function and its width agree well with the star formation rate--halo mass
relation obtained by the abundance matching study of \citet{Bethermin2012}.

In Fig.~\ref{fig:mvir_csmd} we show the colour--stellar mass diagram produced
using the best parameters of our static \Mvir{} model. The black dashed line
indicates the colour split used to divide the galaxies into red and blue
populations.  Although there is a lack of a clear colour bi-modality as seen
in observational data \citep[e.g.][]{Baldry2004}, we still find a clear
overabundance of galaxies with $B{-}V{\approx}0.87$ corresponding to the
observed ``red sequence''.  The presence of this feature at approximately the
correct position in colour space \citep{Cole2005} is an interesting result for
such a simple model.

In the left hand panel of Fig.~\ref{fig:z0_smf} we show the red and blue model
galaxy \SMF{}s (solid lines) against the corresponding constraining observations
(error bars).  Despite its simplicity, the chosen form of the physics function
produces a good reproduction of the data.  This is true across a wide range in
stellar mass, indicating that the model is capable of successfully matching the
integrated time evolution of stellar mass growth as a function of halo mass
at \z0{}.  Also, since blue galaxies preferentially trace those objects which
have undergone significant recent star formation, the model's reproduction of
the observed blue mass function suggests that the rate of star formation as a
function of stellar mass near \z0{} is also in broad agreement with the
observed Universe.

The fact that such an agreement is attainable with this simple model should
be viewed as a key success of the methodology and a validation of the general
form we have chosen for the physics function, \FphysMvir{}.  Having said this,
there are some differences in the left hand panel of Fig.~\ref{fig:z0_smf} worth
noting.  In particular, there is an over-prediction in the number density of
the most massive red galaxies and a corresponding under prediction of the most
massive blue galaxies.

\subsubsection{The \Vmax{} model}
\label{sec:z0_vmax}

Having established that a physics function constructed using \Mvir{} as the
single input variable can successfully provide a good match to the observed
\z0{} red and blue \SMF{}s, we now turn our attention to the results of using
\Vmax{} as the input property (Eqn.~\ref{eqn:physicsfunc_vmax}).  

In the right hand panel of Fig.~\ref{fig:z0_smf} we present the colour-split
\SMF{}s for the static \Vmax{} model.  Again, a fixed colour division of
$B{-}V{=}0.8$ is used to define the two colour populations and we use
MCMC tools to constrain the physics function parameter values to provide
the best statistical reproduction of the \citet{Bell2003} data. The
resulting parameter values are presented in Table~\ref{tab:params}.
Unsurprisingly, a comparison with the equivalent values of the \Mvir{} model
indicates that the peak efficiency of converting fresh baryonic material
into stars in a single time-step remains similar ($\mathcal{E}_{V_{\rm
max}}{=}0.53;\,\mathcal{E}_{M_{\rm vir}}{=}0.56$).  However, the average
virial mass of haloes with $V_{\rm max}{\approx} (V_{\rm peak}{=}158\:
\rm{km\,s^{-1}})$ is $10^{11.9}\,\rm{M_{\sun}}$, therefore this peak efficiency
occurs in slightly more massive haloes than was the case for the \Mvir{}
model.  This is a reflection of the different growth histories of these two halo
properties.

As was the case for the \Mvir{} model, an excellent reproduction of the
observations is attainable when using \Vmax{} as the input parameter to the
physics function.  We find that the over prediction of high-mass red galaxies
has been alleviated, although at the cost of now somewhat under predicting
the number density of low-mass blue galaxies. Importantly though, given a
suitable choice for values of the free parameters of the physics function,
both the \Mvir{} and \Vmax{} physics functions can produce a good match to
the distribution and late time growth of stellar mass at \z0{} despite the
differences in their mean time evolution (cf. Fig.~\ref{fig:cartoon}).

In Fig.~\ref{fig:z0_smf_probs} we present the marginalized posterior probability
distributions for our MCMC calibration of both the \Mvir{} (left-hand panel) and
\Vmax{} (right-hand panel) models.  For clarity we have zoomed in on the regions of
high probability in all panels instead of showing the full ranges explored.
The well behaved and understandable nature of the parameter distributions
gives us further confidence in the validity of our model implementation.  The
approximately Gaussian shape of the 1D distributions (diagonal panels), coupled
with their uni-modal nature, indicates that all of the parameters are well
constrained.  Furthermore, the 2D panels demonstrate that the only degeneracies
in either model are between those parameters controlling the normalization
(\EMvir{}/\EVmax{}) and width (\sigmaMvir{}/\sigmaVmax{}) of the log-normal
physics function.  This makes intuitive sense as these parameters jointly
determine the integral of the star formation rate defined by Eqn.~\ref{eqn:sfr}
and therefore the approximate total amount of stellar mass formed by each
galaxy.

\subsection{High redshift}
\label{sec:highz_results}

\begin{figure}
    \includegraphics[width=\columnwidth]{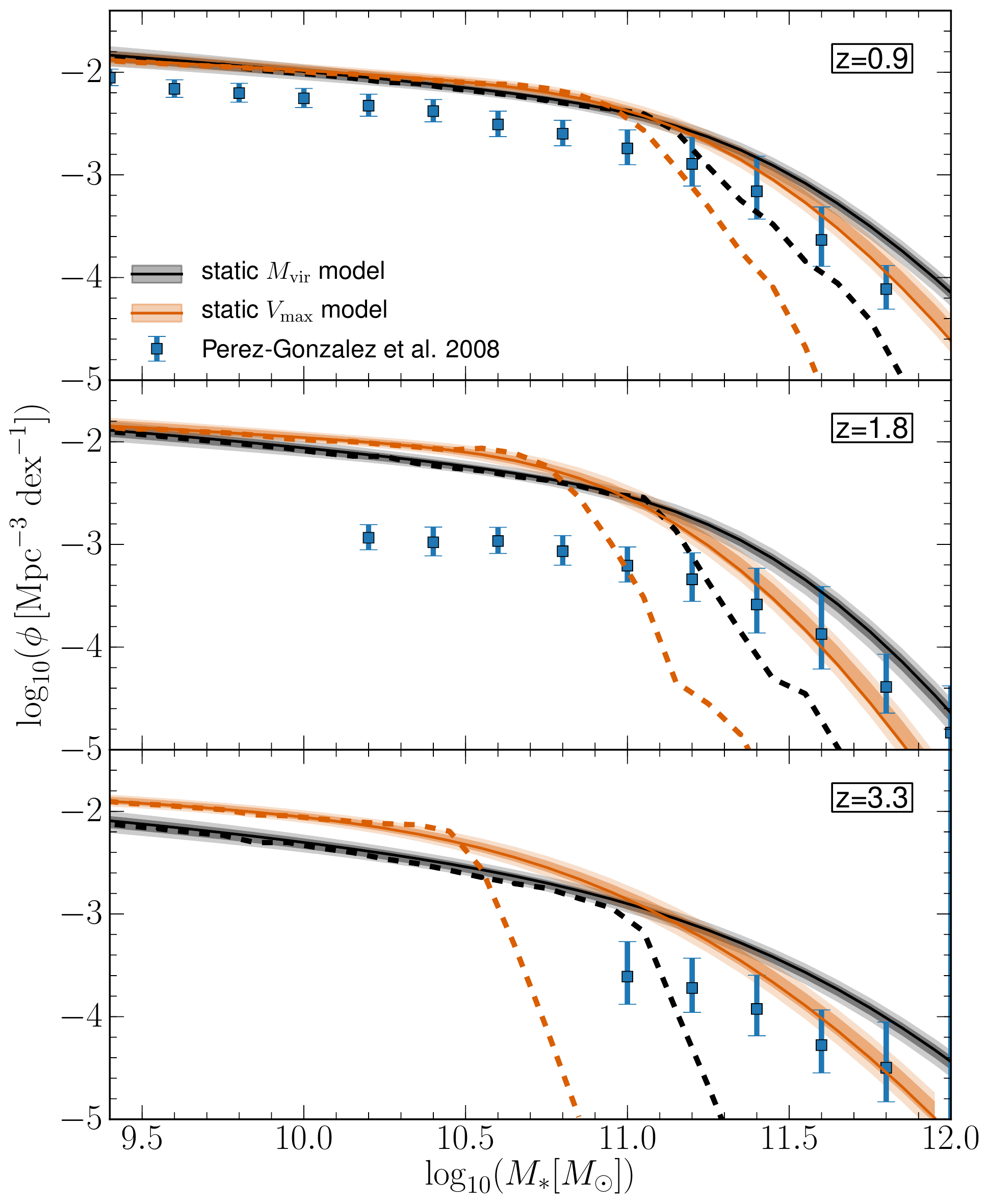} 
    \caption{\label{fig:3panel_smf} The $z{\approx}0.9$,
    1.8, 3.3 \SMF{}s predicted by the static \Mvir{} (dashed
    black line) and \Vmax{} (dashed orange line) models
    (Eqns.~\ref{eqn:physicsfunc_mvir},\ref{eqn:physicsfunc_vmax}).
    Observational data from \citet[][PG08]{Perez-Gonzalez2008} and are shown for
    comparison.  The solid lines indicate the results of convolving the model
    stellar masses with a normally distributed random uncertainty of 0.3 or 0.45
    dex (for redshifts less than/greater than 3, respectively) in order to mimic
    the systematic uncertainties associated with the observed masses.  The dark
    and light shaded regions show the 68 and 95\% confidence intervals predicted
    using the marginalized parameter distributions obtained from our $z{=}0$
    parameter calibration (Fig.~\ref{fig:z0_smf_probs}).  There are clear
    differences between the \Mvir{} and \Vmax{} models at higher redshifts,
    despite their approximate agreement at \z0{} (Fig.~\ref{fig:z0_smf}).  This
    reflects the different time evolution of these two halo properties (see
    Fig.~\ref{fig:cartoon}).}
\end{figure}

In the previous section we demonstrated that our simple formation history
model is capable of reproducing the observed red and blue galaxy \SMF{}s
of the local Universe.  We also showed that this is true independent of
whether we utilize the \Mvir{} or \Vmax{} form of the physics function
(Eqns.~\ref{eqn:physicsfunc_mvir} \& \ref{eqn:physicsfunc_vmax}).  However,
as seen in Fig.~\ref{fig:cartoon}, there are important differences in the
time evolution of these halo properties.  This suggests that we should see
corresponding differences in the galaxy populations predicted at higher
redshifts.

In Fig.~\ref{fig:3panel_smf} we present the \SMF{}s of both the \Mvir{} and
\Vmax{} models (dashed lines) against the observed $z{>}0$ mass functions of
\citet[][points]{Perez-Gonzalez2008}.  The solid lines represent the formation
history model results after a convolution with a normally distributed random
error of dispersion 0.3 dex for $z{<}3$ and 0.45 dex for redshifts greater
than this value \citep{Moster2013}.  Such a convolution is common practice
and approximates the missing uncertainties in the observational data due to
systematics involved with producing stellar mass estimates from high-redshift
galaxy observables \citep[e.g.][]{Fontanot2009,Guo2011,Santini2012}.

There are clear quantitative differences between the stellar mass functions
produced by the two models.  These become more pronounced as we move to higher
redshifts.  At $z{\approx}3$ (bottom panel), the \Vmax{} model predicts a sharp
fall off in the number density of galaxies with stellar masses greater than
$10^{10.5}\,M_{\sun}$.  When using \Mvir{} to define the physics function,
this drop off does not occur until $M_{\star}{\approx} 10^{11}\,M_{\sun}$,
resulting in a differing prediction in the number density of these galaxies
by greater than two orders of magnitude at high masses.  Despite this, both
versions of the physics function predict $z{>}0$ \SMF{}s which are too steep
at high masses, although the addition of the random uncertainties (solid
lines) largely alleviates this problem.  There are also notable differences at
lower stellar masses, where both models over-predict the number of galaxies.
The \Vmax{} model also predicts that a large fraction of galaxies with
$M_{\rm vir}{<}10^{10.5}\,M_{\sun}$ are already in place by $z{=}3$, with a
correspondingly slower evolution to $z{=}0$.

Many of these differing qualitative predictions can be understood by considering
the differences in the time evolution of \Mvir{} and \Vmax{} as shown in
Fig.~\ref{fig:cartoon}.  For example, the deficit of high stellar mass galaxies
in the \Vmax{} model at $z{\approx}3$ is due to their host haloes initially being
identified with \Vmax{} values greater than \VmaxPeak{} (at least for the mass
resolution of our simulation).  These haloes therefore spend little time in
the efficient star-forming band.  The result is a reduced amount of in situ
star formation at early times, with effects that carry all the way through to
\z0{} as these galaxies grow, predominantly through merging.  We can similarly
understand the cause of the larger predicted number density of high-redshift
low-mass galaxies in the \Vmax{} model.  In this case, the lowest mass haloes present
at high redshifts have spent a longer time close to the peak of the efficient
star-forming band.  This results in these haloes already hosting significant
amounts of stellar mass by $z{=}3$.

\begin{figure*}
  \begin{minipage}{\textwidth}
    \begin{center}
      \includegraphics[width=1.0\textwidth]{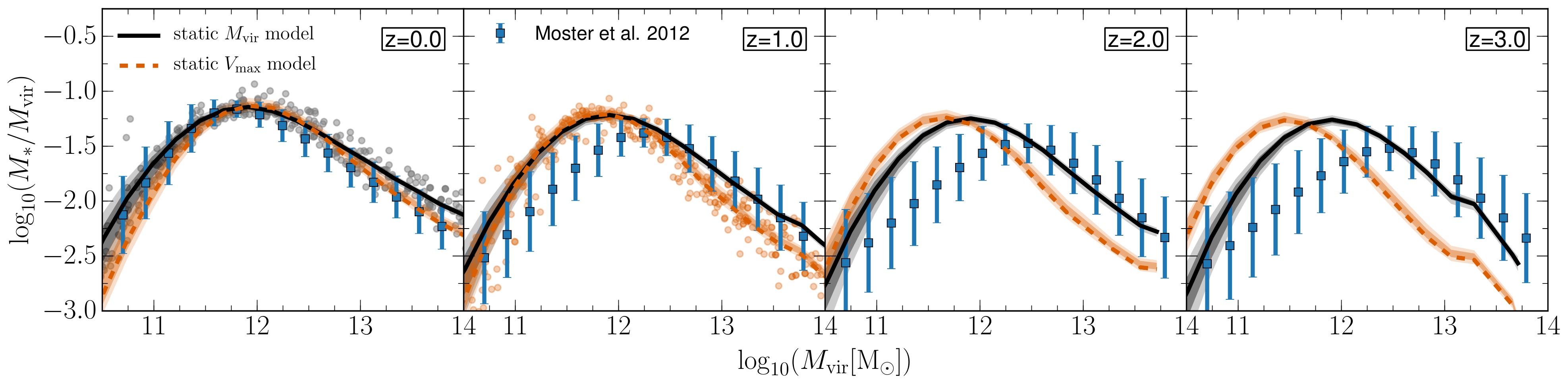}
      \caption{\label{fig:no_evo_SHMrelation} The evolution of the mean
      stellar--halo mass relation of central galaxies. The results of
      the static \Mvir{} (Eqn.~\ref{eqn:physicsfunc_mvir}; black solid
      line) and \Vmax{} (Eqn.~\ref{eqn:physicsfunc_vmax}; orange dashed
      line) models are both shown. The dark and light shaded regions
      indicate the 68 and 95\% confidence intervals predicted using the
      marginalized parameter distributions from our calibration at $z{=}0$ (see
      Fig.~\ref{fig:z0_smf_probs}).  A comparison with the subhalo abundance
      matching results of \citet{Moster2013} (blue error bars) indicates that
      both forms of the physics function fail to reproduce the evolution of the
      stellar mass growth efficiency required to reproduce the high-redshift
      stellar mass function (see Fig.~\ref{fig:3panel_smf}).  Coloured dots in
      the first and second panels show a random sampling of 20 galaxy--halo
      systems from each $x$-axis bin of the \Mvir{} and \Vmax{} models,
      respectively.}
    \end{center}
  \end{minipage}
\end{figure*}

To illustrate this further, in Fig.~\ref{fig:no_evo_SHMrelation} we show
the evolution of the mean stellar--halo mass relation for both models. The
blue error bars represent the relations predicted by the subhalo abundance
matching model of \citet{Moster2013}. We have specifically chosen to compare
our results against the work of \citet{Moster2013}, as they take their halo
masses from the same dark matter merger trees as used in this work \citep[as
well as the higher resolution Millennium-II Simulation;][]{Boylan-Kolchin2009}
and also construct their model to match the same high-redshift stellar mass
functions of \citet{Perez-Gonzalez2008}. Hence, the blue error bars of
Fig.~\ref{fig:no_evo_SHMrelation} represent the evolution in the integrated
stellar mass growth efficiency which our model must achieve in order to
successfully replicate the observed \SMF{}s of Fig.~\ref{fig:3panel_smf}.

By construction, both the \Mvir{} and \Vmax{} models produce extremely similar
relations at $z{=}0$ but with clear differences at higher redshifts. It
is these variations in the typical amount of stars formed within haloes of
a given mass that drives the different predictions for the evolution of
the stellar mass function. For example, the much higher average stellar
mass content of low-mass haloes at $z{=}3$ when using the \Vmax{} model
(Fig.~\ref{fig:no_evo_SHMrelation}) is the cause of the increased normalization
of the low-mass end of the relevant \SMF{} in Fig.~\ref{fig:3panel_smf}.

Importantly, it can be seen from Fig.~\ref{fig:no_evo_SHMrelation} that neither
the \Mvir{} nor the \Vmax{} model reproduces the evolution of the stellar--halo
mass relation found by \citet{Moster2013}; in particular the position and
normalization of the peak value. The use of a redshift-independent halo mass
to define the peak in situ star formation efficiency of the \Mvir{} model
results in no change to the position of the peak of the stellar--halo mass
relation with redshift.  Although not immediately obvious why this should be
so, it can be understood by considering the typical evolution of a halo
across the relation.

At early times, haloes grow in mass rapidly, however, they typically still sit
below the efficient star formation mass regime defined by the physics function
(see Fig.~\ref{fig:cartoon}).  In Fig.~\ref{fig:no_evo_SHMrelation}, these
haloes will therefore travel almost horizontally from left to right with a low
stellar--halo mass fraction.  Eventually haloes will enter the mass regime of
efficient star formation, causing them to rapidly increase their stellar--halo
mass fractions with only a relatively modest growth in halo mass.  This phase
of rapid stellar mass growth causes a ''pile-up'' of galaxies in the \SHMR{}
that peaks around the virial mass at which haloes again transition out of the
efficient star forming regime.  Since the mass at which this occurs is fixed
in our simple static model, the position of the \SHMR{} peak is therefore also
fixed in the \Mvir{} model.  Due to the evolving \Mvir{}--\Vmax{} relationship,
the position of the peak efficiency for the \Vmax{} model does evolve, but
unfortunately in the direction opposite to that required. The shallower tail of
the relation towards higher halo masses is caused by the subsequent growth of
galaxies due to mergers.

As well as the precise shape and normalization of the mean \SHMR{}, it is
also important to consider the scatter of the distribution about this mean.
For example, at high halo masses it is possible to increase the scatter of
stellar--halo mass ratios to produce an increased normalization of the high-mass
end of the stellar mass function whilst leaving the mean \SHMR{} unchanged.
In the first two panels of Fig.~\ref{fig:no_evo_SHMrelation} we have plotted
the stellar--halo mass ratios of 20 randomly selected haloes from each of
the 15 mass bins used to construct the mean relations.  At halo masses above
the peak value we find an approximately constant value for the scatter as a
function \Mvir{} in both models.  Below the peak halo mass, the scatter rapidly
increases with decreasing \Mvir{}.  This reflects the stochastic nature of star
formation for lower mass haloes whose mass evolution may not be well resolved
at all times in our model.  However, at \z0{} we find an average 1$\sigma$
scatter of 0.15 for the \Mvir{} model and 0.23 for the \Vmax{} model over
the range of halo masses plotted.  This agrees well with previous studies
\citep[e.g.][]{More2009,Yang2009,Behroozi2013b}.

Based purely on the inability to reproduce the required evolution in the
stellar--halo mass relation, it is unlikely that the non-evolving physics
function will be able to match the observed distribution of stellar masses
in both the low- and high-redshift Universe simultaneously.  This is true
irrespective of the values of the available parameters or whether \Mvir{} or
\Vmax{} is used as the dependant variable.

\subsection{Incorporating a redshift evolution}
\label{sec:evo_results}

\begin{figure*}
  \begin{minipage}{\textwidth}
    \begin{center}
      \subfigure{\includegraphics[width=0.475\textwidth]{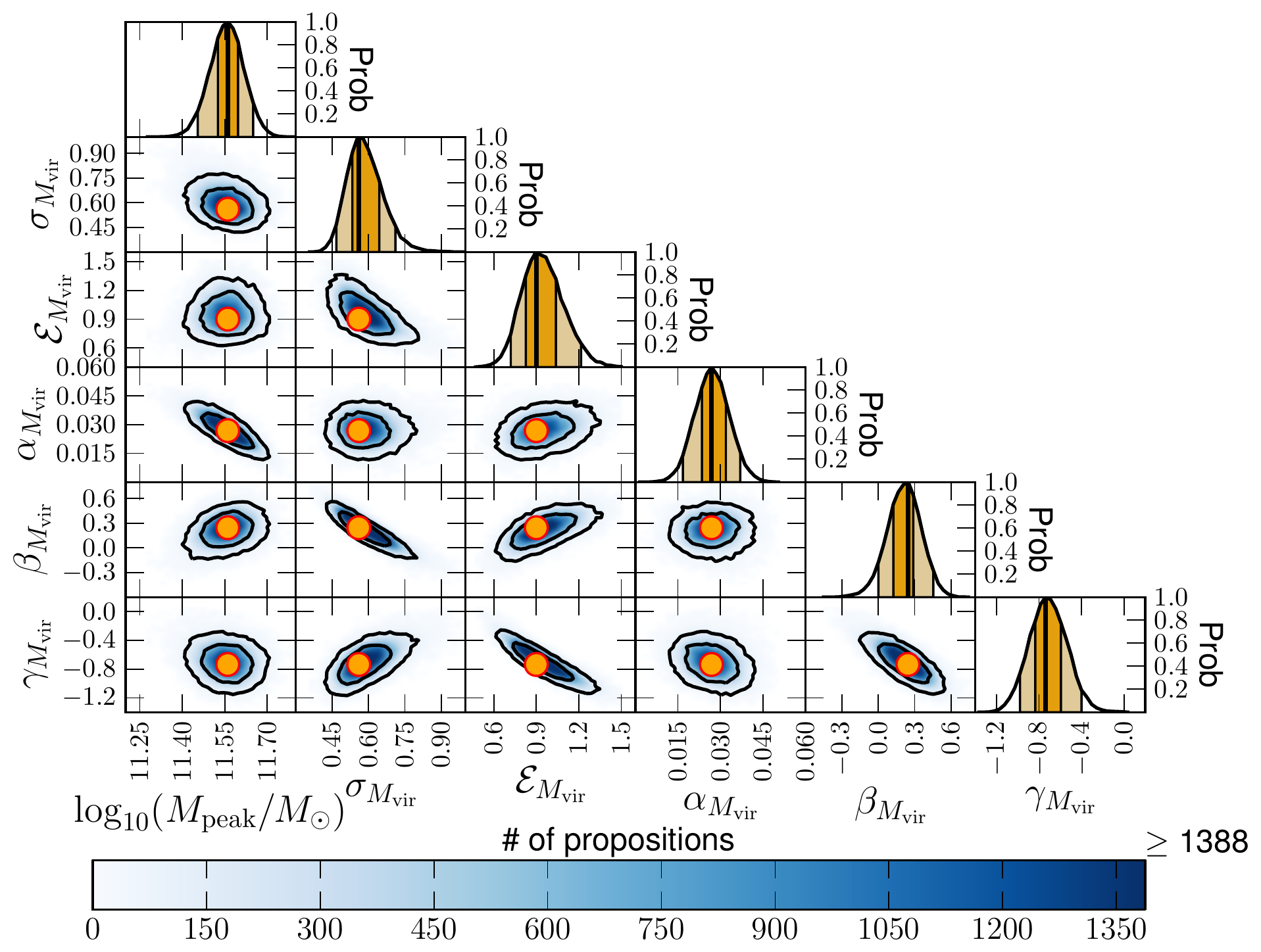}} \quad
      \subfigure{\includegraphics[width=0.475\textwidth]{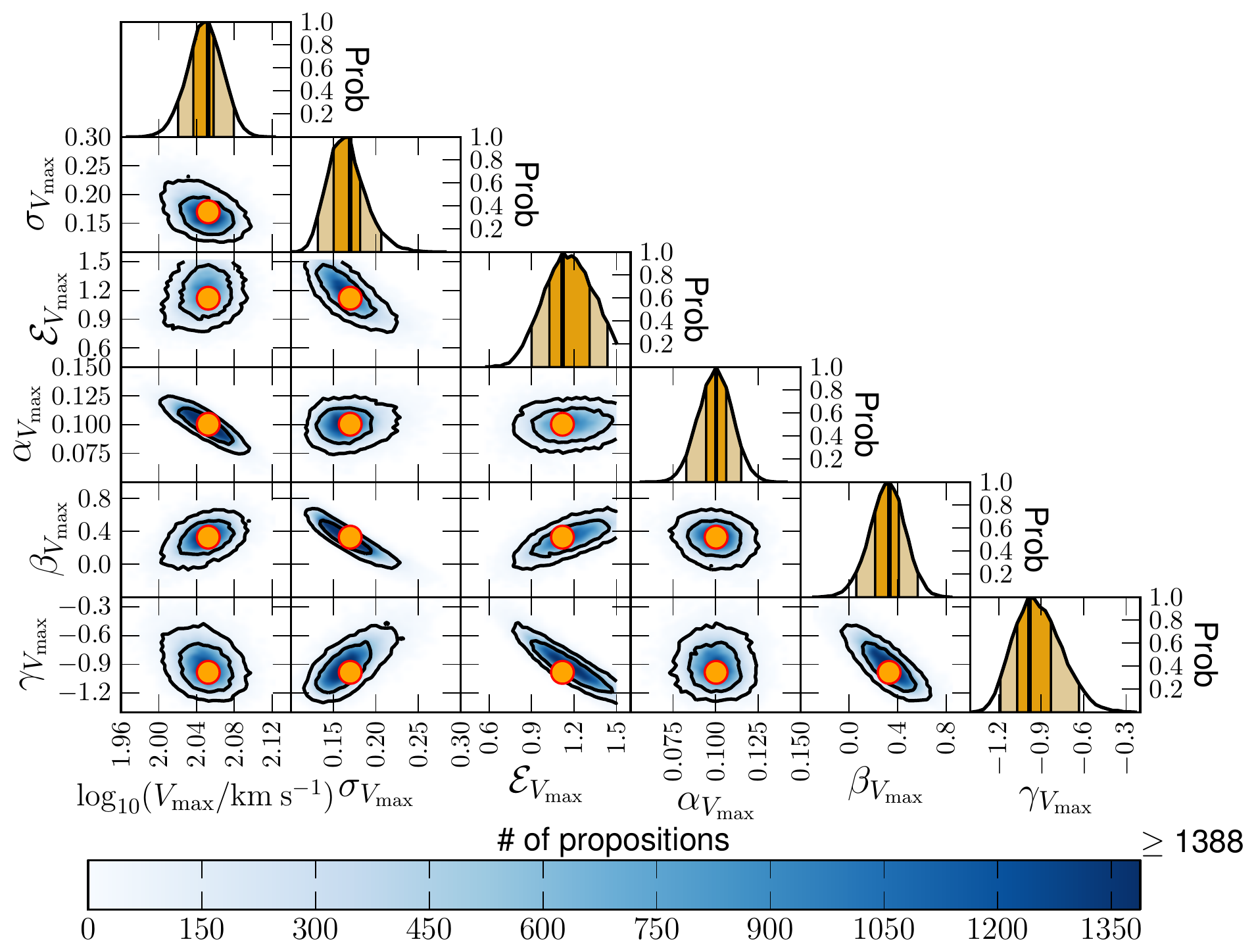}}
      \caption{\label{fig:evo_prob_join} Marginalized posterior probability
        distributions for the \Mvir{} (left) and \Vmax{} (right) parameters of
        the redshift dependent physics function.  In both panels, the models were
        constrained to simultaneously reproduce the observed $z{=}0$ red and blue
        \SMF{}s (Fig.~\ref{fig:evo_colorsplit_SMF}) as well as the time evolution
        of the \SHMR{} (Fig.~\ref{fig:evo_prob_join}).  All panels have been
        zoomed in to the high-probability regions.  Con tours on the 2D (blue)
        panels indicate the 68 and 95\% confidence regions.  Yellow dots mark the
        marginalized most probable parameter values.  The diagonal panels show
        the marginalized 1D distributions with the 1$\sigma$ and 2$\sigma$ confidence
        intervals shown by light and dark shaded regions, respectively.  The
        approximately Gaussian shape of these 1D distributions indicates
        the well behaved nature of the model.}
    \end{center}
  \end{minipage}
\end{figure*}

\begin{figure*}
  \begin{minipage}{\textwidth}
    \begin{center}
      \includegraphics[width=1.0\textwidth]{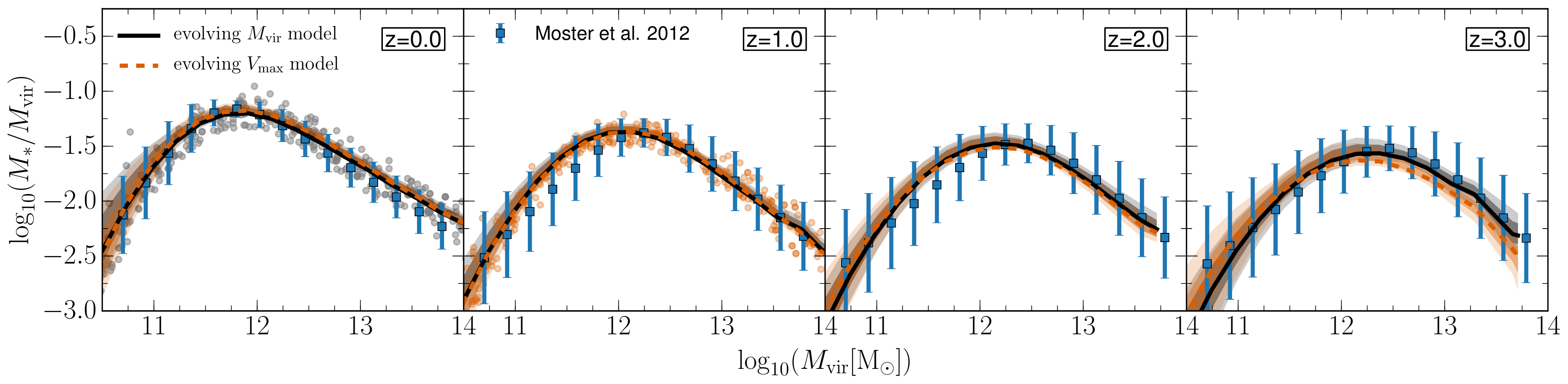}
      \caption{\label{fig:evo_SHMrelation} 
      The evolution of the mean stellar--halo mass relation of central galaxies
      for the evolving \Mvir{} (Eqn.~\ref{eqn:physicsfunc_mvir}; black
      solid line) and \Vmax{} (Eqn.~\ref{eqn:physicsfunc_vmax}; grey
      dashed line) models. Orange shaded regions indicated the subhalo
      abundance matching results of \citet{Moster2013}.  A comparison with
      Fig.~\ref{fig:no_evo_SHMrelation} indicates that by shifting both the
      normalization (\EMvir{}, \EVmax{}) and position (\MvirPeak{}, \VmaxPeak{})
      of the physics functions, we are able to reproduce the correct evolution
      of the stellar--halo mass relation at high-redshifts in terms of both
      shape and amplitude.  This leads to a much better agreement between
      the observed and predicted stellar mass functions at $z{>}0$. (see
      Fig.~\ref{fig:evo_3panel_smf}).  Coloured dots in the first and second
      panels show a random sampling of 20 galaxy/halo systems from each x-axis
      bin of the \Mvir{} and \Vmax{} models respectively.}
    \end{center}
  \end{minipage}
\end{figure*}

Although capable of reproducing the observed red and blue stellar mass
functions at \z0{}, we showed in \S\ref{sec:highz_results} that our
simple formation history model struggles to reproduce the high-redshift
distribution of stellar masses.  Importantly, we also concluded that there
is unlikely to be any combination of physics function parameter values
(see Eqns.~\ref{eqn:physicsfunc_mvir} \& \ref{eqn:physicsfunc_vmax}) which
could alleviate this discrepancy.  In this section we therefore look to
extend our simple model by introducing a redshift dependence to the physics
function.  This is equivalent to the introduction of an evolution
of the star formation efficiency with time for a fixed halo mass/maximum
circular velocity.  Such an evolution is well motivated both theoretically
and observationally, suggesting the presence of alternative/additional
star formation mechanisms at high-redshift when compared to those of
the local Universe.  For example, so-called ``cold-mode'' accretion
\citep{Birnboim2003,Keres2005,Brooks2009} is thought to be able to effectively
fuel galaxies of massive haloes at high-redshift, allowing for increased star
formation.  In addition, the early Universe was a more dynamic place with
an enhanced prevalence of gas-rich galaxy mergers and turbulence-driven star
formation \citep[e.g.][]{Dekel2009b,Wisnioski2011}.

To reproduce the evolving position and normalization of the stellar--halo mass
relation as found by \citet{Moster2013}, we modify the physics function of
Eqn.~\ref{eqn:physicsfunc_mvir} by introducing a simple power law dependence on
redshift to each of the free parameters:
\begin{eqnarray}
  \log_{10}(M_{\rm peak}(z)) &=& \log_{10}(M_{\rm peak})
  (1+z)^{\alpha_{M_{\rm vir}}}\ ,\\
  \sigma_{M_{\rm vir}}(z) &=& \sigma_{M_{\rm vir}}
  (1+z)^{\beta_{M_{\rm vir}}}\ ,\\
  \mathcal{E}_{M_{\rm vir}}(z) &=& \mathcal{E}_{M_{\rm vir}}
  (1+z)^{\gamma_{M_{\rm vir}}}\ ,
\end{eqnarray}
where at $z{=}0$: \MvirPeak{}$(z){=}$\MvirPeak{},
\sigmaMvir{}$(z){=}$\sigmaMvir{} and \EMvir{}$(z){=}$\EMvir{}.  The exact
values of the redshift scalings are calibrated using MCMC to provide the best
simultaneous reproduction of the \citet{Moster2013} \SHMR{} at $z{=}0$, 1,
2 and 3, as well as the $z{=}0$ red and blue \SMF{}s, and are presented in
Table~\ref{tab:params}.  

In the left hand panel of Fig.~\ref{fig:evo_prob_join} we present the
relevant marginalized posterior probability distributions of the six
free model parameters.  Similarly to the redshift-independent case
(cf. \S\ref{sec:z0_results}), the approximately Gaussian shape of the 1D
probability distributions indicates that the parameters are generally
well constrained.  However, in addition to the degeneracy between the
physics function normalization (\EMvir{}) and width (\sigmaMvir{}) noted in
\S\ref{sec:z0_results}, there are also clear and understandable degeneracies
between the redshift evolution and $z{=}0$ value of each parameter
(e.g. \MvirPeak{} and $\alpha_{M_{\rm vir}}$).

We note that there are minor differences between the \z0{} \SMF{} utilized by
\citet{Moster2013} to constrain their \SHMR{} \citep{Li2009}, and the \z0{} mass
function which we employ in this work \citep{Bell2003}.  However, we calibrate
our model against both the \SHMR{} and colour-split stellar mass functions at
\z0{} with equal weights.  The MCMC fitting procedure then attempts to find
the best compromise between these two (as well as all other) constraints.  Since
we find that there are no multimodal features in the marginalized posterior
probability distributions (see Fig.~\ref{fig:evo_prob_join}), the parameter sets
required to fit each constraint individually must be statistically compatible
with each other.  We therefore conclude that this slight inconsistency in
our fitting procedure has minimal effect on our ability to demonstrate the
success and utility of the model and on our results.

The preferred values of $\alpha_{M_{\rm vir}}$ and $\beta_{M_{\rm vir}}$
are relatively small, indicating little need for evolution in both the peak
position, \MvirPeak{}$(z)$, and width, \sigmaMvir{}$(z)$, of the physics
function.  As a consequence, the values of both \MvirPeak{} and \sigmaMvir{}
are similar to the non-evolving case (cf. Table~\ref{tab:params}).  However,
there is a strong evolution preferred for the normalization of the physics
function, $\gamma_{M_{\rm vir}}$, such that it decreases rapidly with increasing
redshift.  In order to maintain the total $z{=}0$ stellar mass density, the
value of \MvirPeak{}=0.9 is therefore considerably higher than was the case
in the non-evolving form of the physics function.  This implies that 90\%
of all freshly accreted baryonic material in haloes with $\log_{10}(M_{\rm
peak}/M_{\sun}){=}11.6$ is converted into stars at $z{=}0$.  However, at
$z{=}1$,2 and 3 the peak conversion efficiencies are considerably lower: 54,
40 and 32\%, respectively.

We also similarly modify the \Vmax{} physics function, \FphysVmax{}:
\begin{eqnarray}
  \log_{10}(V_{\rm peak}(z)) &=& \log_{10}(V_{\rm peak})
  (1+z)^{\alpha_{V_{\rm max}}}\ ,\\
  \sigma_{V_{\rm max}}(z) &=& \sigma_{V_{\rm max}}
  (1+z)^{\beta_{V_{\rm max}}}\ ,\\
  \mathcal{E}_{V_{\rm max}}(z) &=& \mathcal{E}_{V_{\rm max}}
  (1+z)^{\gamma_{V_{\rm max}}}\ ,
\end{eqnarray}
with the redshift scalings being calibrated to reproduce the same observations
as the \Mvir{} case above.  The marginalized posterior probability distributions
are presented in the right-hand panel of Fig.~\ref{fig:evo_prob_join}, with the
preferred parameter values again presented in Table 1.

As was found for the redshift-dependent \Mvir{} model, the marginalized
posterior distributions indicate that the free model parameters are well
constrained and that there are no unexpected degeneracies between them
(Fig.~\ref{fig:evo_prob_join}).  If we consider the preferred values of the
parameters themselves, we again find that the normalization of the physics
function, \EVmax{}$(z)$, shows the most pronounced evolution.  A value of
$\gamma_{V_{\rm max}}{=}-0.98$ indicates that the maximum star formation
efficiency declines almost linearly as a function of $(1{+}z)$.  This strong
evolution requires a value of \EVmax{} at $z{=}0$ that is actually greater
than 1, and hence more than the total freshly accreted baryonic material
must be converted into stars in haloes with $\log_{10}(V_{\rm peak}/{\rm
(km\,s^{-1})}){=}2.1$ at this redshift.  This suggests that we may need to vary
the universal baryon fraction as a function of halo mass (or maximum circular
velocity), perhaps to mimic the effects of processes such as the recycling of
ejected baryons during star formation \citep[e.g.][]{Papastergis2012}.

\begin{figure}
  \includegraphics[width=\columnwidth]{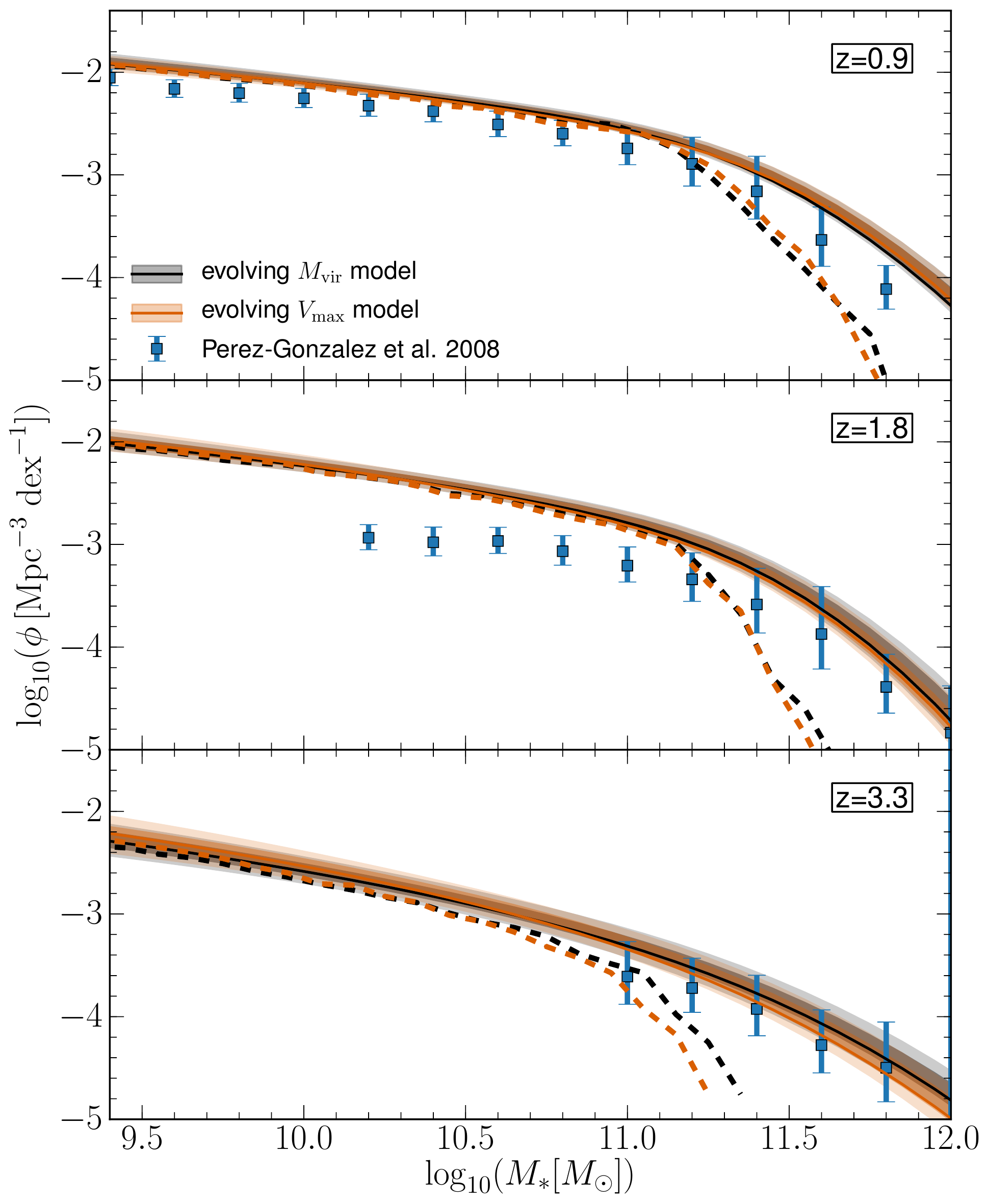}
  \caption{\label{fig:evo_3panel_smf} The $z{\approx}0.9$, 1.8, and 3.3 \SMF{}s
  predicted by the evolving \Mvir{} (dashed black line) and \Vmax{} (dashed
  orange line) models.  Observational data from \citet{Perez-Gonzalez2008} are
  shown for comparison.  The solid lines give the results of convolving the
  model stellar masses with a normally distributed random uncertainty of 0.3 or
  0.45 dex (for redshifts less than/greater than 0.3, respectively) in order to
  mimic the systematic uncertainties associated with the observed masses.  By
  using an appropriate redshift evolution of the physics function parameters,
  the model's ability to successfully recover the observed high-redshift stellar
  mass functions is improved.}
\end{figure}

\begin{figure*}
  \begin{minipage}{\textwidth}
    \begin{center}
      \subfigure{\includegraphics[width=0.475\textwidth]{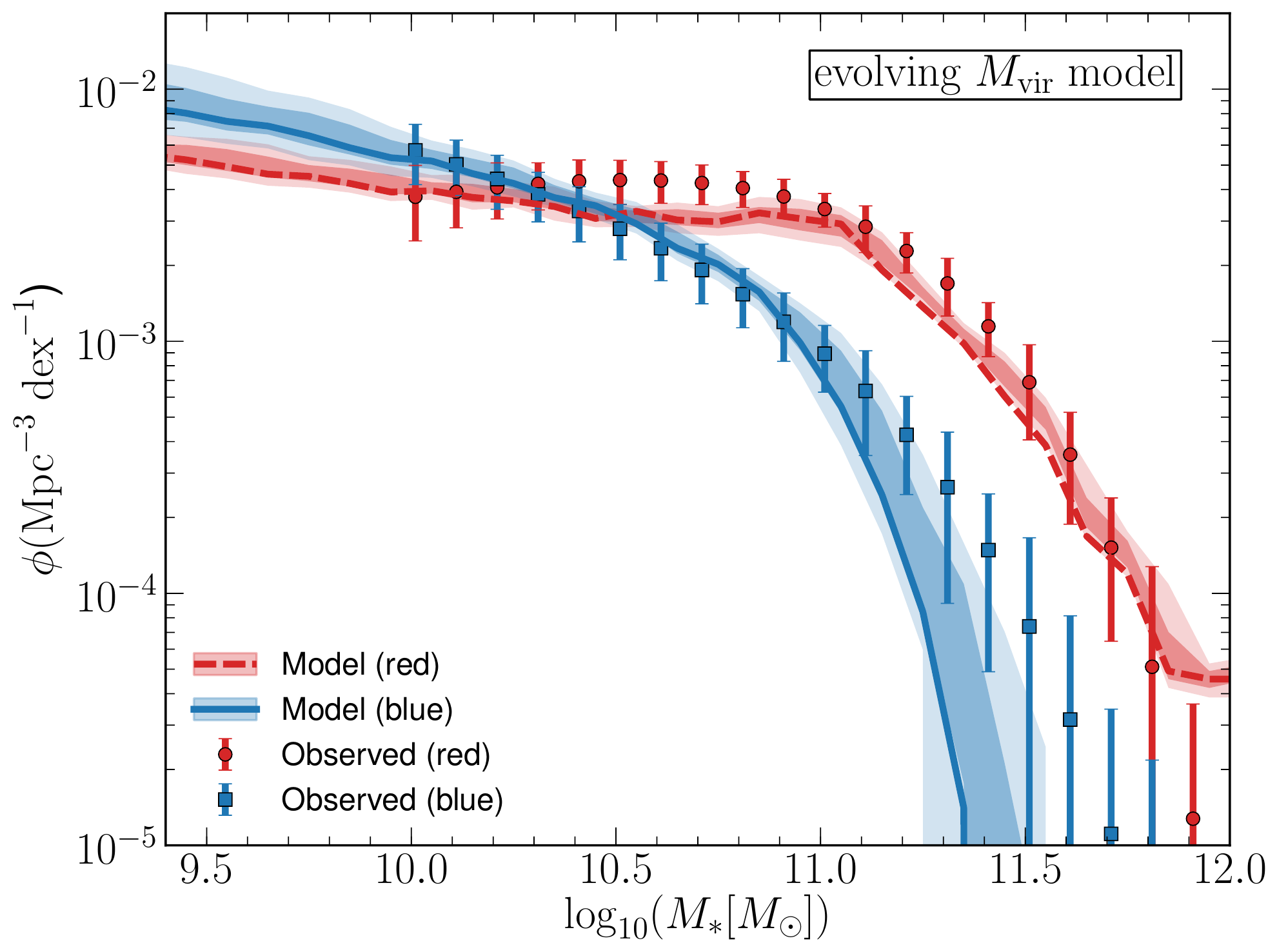}}
      \quad
      \subfigure{\includegraphics[width=0.475\textwidth]{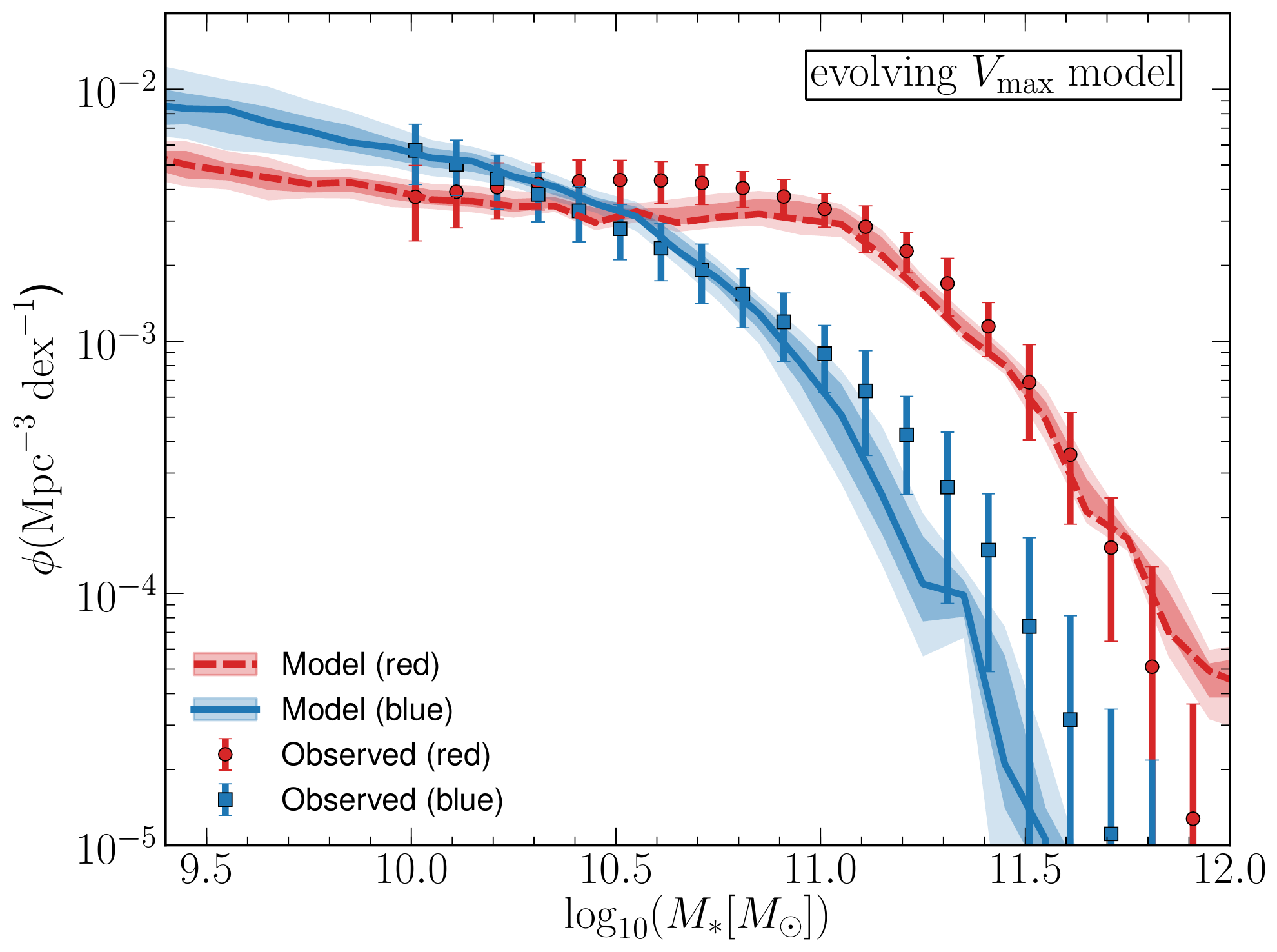}}
      \caption{\label{fig:evo_colorsplit_SMF} The red (dashed lines) and blue
      (solid lines) galaxy \SMF{}s produced by the evolving \Mvir{} (left)
      and \Vmax{} (right) formation history models.  Error bars indicate the
      observations of \citet{Bell2003}.  The free model parameters have been
      constrained to simultaneously reproduce these mass functions as well
      as the evolution of the \SHMR{} (Fig.~\ref{fig:evo_SHMrelation}).  The
      dark and light shaded regions show the associated 68 and 95\% confidence
      regions obtained from this calibration.}
    \end{center}
  \end{minipage}
\end{figure*}

In Fig.~\ref{fig:evo_SHMrelation}, we present the stellar--halo mass relations
of the new, redshift-dependant, \Mvir{} (black) and \Vmax{} (orange) models.
The blue error bars again indicate the results of \citet{Moster2013}.  By
incorporating the redshift dependence we are now able to successfully reproduce
the evolution of both the normalization and peak position of the stellar--halo
mass relation required at $z{\ge}0$.  The effects of this on the predicted
high-redshift stellar mass functions of both the \Mvir{} and \Vmax{} models
can be seen in Fig.~\ref{fig:evo_3panel_smf}.  As expected, we now find an
improved agreement with the observations when compared to the original,
non-evolving physics function results (cf. Fig.~\ref{fig:3panel_smf}).  The
typical 1$\sigma$ scatter in the evolving \Mvir{} \SHMR{} remains unchanged
from the static case at approximately 0.15 dex at \z0{}.  However, the scatter
in the evolving \Vmax{} model is reduced to approximately 0.19 dex (from 0.23
dex in the static case).  In both models the scatter decreases as a function of
redshift such that at $z{=}1$ and $2$ it is approximately 0.13 and 0.01 dex,
respectively.

For completeness we also present the \z0{} colour-split \SMF{}s for both
models in Fig.~\ref{fig:evo_colorsplit_SMF}.  A reasonable agreement with the
constraining observational data is still achieved.  However, we now find that an
underprediction in the number density of massive blue galaxies is present in
both models, suggesting that our implemented evolutionary model may not provide
enough late time star formation in the most massive haloes.

\section{Discussion}
\label{sec:discussion}

In \S\ref{sec:z0_results}, we demonstrated that our most basic, non-evolving
form of the physics function is able to successfully reproduce the observed red
and blue stellar mass functions of the local Universe (Fig.~\ref{fig:z0_smf}).
This key result highlights the utility and validity of our basic methodology and
model implementation.  In addition, it reinforces the commonly held belief that
the growth of galaxies is intrinsically linked to the growth of their host dark
matter haloes \citep{White1978}.

Although the level of agreement achieved with the observed $z{=}0$ colour-split
\SMF{}s is generally very good, there are some discrepancies.  For example,
there is an underprediction in the number density of the most massive blue
galaxies in the \Mvir{} model (left-hand panel of Fig.~\ref{fig:z0_smf}), with
a corresponding overprediction in the number of massive red galaxies.  Our
$z{>}0$ analysis suggests that this is at least partially due to an incorrect
evolution of the \SHMR{} with time (see Fig.~\ref{fig:no_evo_SHMrelation}).
However, we also note that an excess of massive red galaxies is a common
feature of traditional semi-analytic galaxy formation models which similarly
tie the evolution of galaxies to the masses of their host dark matter haloes.
In such models, efficient feedback from AGN is typically responsible for
truncating star formation in the most massive galaxies and hence causes the
average stellar populations of these objects to become older and redder
\citep{Bower2006,Croton2006,Mutch2011}.  This is already mimicked within the
framework of our formation history model through the turnover at the high
\Mvir{} (or \Vmax{}) end of the physics function.  However, a more gradual
cut-off may be required in the \Mvir{} model case, in order to allow star
formation to proceed for longer in the galaxies populating the most massive
haloes.

In this paper we have deliberately restricted ourselves to considering only a
very simple form of the physics function.  This has allowed us to take advantage
of the resulting transparency when interpreting our findings. However, we stress
that the model can easily be extended to include arbitrary levels of complexity.
For example, we have chosen to use a log-normal distribution to define the form
of the physics function.  Although being conceptually simple, the symmetric
nature of this formalism implicitly assumes that the physical mechanisms
responsible for quenching star formation in both low- and high-mass haloes scale
identically with halo mass (Fig.~\ref{fig:cartoon}).  This assumption has little
physical justification and in order to provide the best results, it may be
necessary to independently adjust the slope of the function at both low and high
masses, and perhaps even as a function of redshift.  In future work, we will
address this issue by carrying out a full statistical analysis aimed at testing
a number of different functional forms for both the physics and baryonic growth
functions.

Even within the reduced scope of this current work, we have learned a great deal
from simply examining the high-redshift stellar mass function predictions of the
formation history model.  In particular, we have highlighted the need for the
physics function to produce an evolution in the stellar--halo mass relation as
a function of redshift in order to match the observed space density of massive
galaxies at early times.  Using \Vmax{} as the input property to the function
introduces such an evolution, but in the wrong direction.  Future improvements
to the model could focus on finding a halo property that does evolve correctly
with time and would thus be a more natural anchor of the physics function.
This would avoid the need to artificially introduce an evolution to match the
observations, as we have done here.

Although the need for an evolving stellar--halo mass relation has been discussed
in the literature, the precise form with which this evolution manifests itself
is less clear.  The results of subhalo abundance matching studies, such as
that of \citet{Moster2013} (which we compare to in this work), are quite
sensitive to the choice of input data sets and the technical aspects of the
methodology.  For example, \citet{Moster2013} find that the peak stellar--halo
mass ratio increases from just 0.15\% at ${z=}4$ to 4\% at $z{=}0$, with a
corresponding shift in position from a halo mass of $10^{12.5}\,{\rm M_{\sun}}$
to $10^{11.8}\,{\rm M_{\sun}}$.  In contrast, an alternative study carried out
by \citet{Behroozi2013b} finds very little change in either the normalization
or peak location over a broad range in redshift.  However, they do note a
marked drop in the relation for the most massive haloes at $z{\la}2.5$.  This
results in a qualitatively different prediction for the evolution of these
massive haloes, such that their efficiency of converting baryons in to stars is
higher as a function of look-back time \citep[the opposite trend to that found
by][]{Moster2013}.

A potentially valuable use for the formation history model is to provide a
general consistency check of subhalo abundance matching studies.  The physics
function could be adapted to exactly replicate the shape and evolution of
the star formation efficiencies they predict \citep{Behroozi2013}, allowing
their validity to be assessed when self-consistently applied to individual
dark matter merger trees.  The additional galaxy properties provided by the
formation history model, such as star formation histories and colours, could
be used to further compare and contrast the success of different abundance
matching methodologies.  For example, it has been suggested that galaxy mergers
may result in a significant fraction of the in-falling satellite stellar mass
being added to a diffuse ICL component instead of to the newly formed galaxy
\citep[][]{Monaco2006,Conroy2007}.  The strength of this effect is expected to
increase significantly with increasing halo mass and is included in the subhalo
abundance matching study of \citet{Behroozi2013b} but not \citet{Moster2013}.
By simply adding a mass-dependent amount of stellar material to an ICL component
during merger events, our simple formation history model could be easily adapted
to explore such a scenario.

We also note that the simplicity of our formation history model results in
it being extremely fast and computationally inexpensive when compared to
traditional semi-analytic models.  This has allowed us to straightforwardly
calibrate it against a number of observed relations using MCMC techniques.  This
procedure can also be trivially extended to provide statistically accurate
(against select observations) mock catalogues for use with large surveys.
Further to what can be achieved using current HOD or subhalo abundance matching
methods, catalogues produced using our model include both full growth histories
and star formation rate information for each individual galaxy, with no need to
add any artificial scatter to approximate variations in formation histories.  In
addition, the direct and clear dependence of the model on the halo properties
of the input dark matter merger trees makes it an ideal tool for investigating
a number of additional topics.  Examples include comparing the effects of
variations between different $N$-body simulations and halo finders on the
physics of galaxy formation and evolution, investigating the predictions of
simple monolithic collapse scenarios, contrasting various mass-dependent merger
starburst models and exploring the ramifications of $N$-body simulations run
with alternative theories of gravity.

Finally, we note that a common criticism of semi-analytic models is the
presence complex degeneracies between large numbers of free parameters.  The
MCMC calibration procedure we employ highlights the complete absence of such
degeneracies in our formation history model, further demonstrating its well
behaved and understandable nature.

\subsection{Potential model extensions}

One area of the model presented in this work which may benefit from being
extended is the treatment of mergers (cf. \S\ref{sec:generating_galaxy_pop}).
In the current model, merger-driven starbursts occur immediately when an
infalling satellite halo crosses the virial radius of its parent.  All of the
newly formed stars are then added to the central galaxy of the parent halo.
However, it is likely that there will be a significant time delay between
the satellite crossing the virial radius of the parent and the actual merger
between the satellite and central galaxy. In practice, due to the relatively
large temporal spacing of our input dark matter merger trees (${\approx}
200{-}350\,{\rm Myr}$), we expect this slight inconsistency to have little effect
on our results.  However, if running the formation history model on merger
trees with a higher temporal resolution, this issue may become important.  A
future investigation of the clustering predictions of the formation history
model, especially when split by galaxy colour, will allow us to fully assess the
validity of these simplifications.

Also, as discussed in \S\ref{sec:generating_galaxy_pop}, we assume that all
freshly accreted baryonic material is available for star formation, regardless
as to whether or not it is already locked up in stars in the form of an
infalling satellite galaxy.  In practice this simply leads to an increased
star formation efficiency for merger events.  Testing of an alternative model
in which the d\Mvir{} term of the baryonic growth function includes only
smooth accretion (i.e. does not include mass increases due to the accretion of
satellite haloes) and star formation is allowed to proceed in satellite galaxies,
indicates that merger-driven starbursts are an important feature of our model.
Without this efficient star formation mechanism there is no combination of the
available free parameters that allows the static model to reproduce the local
colour-split stellar mass function.

Another potential extension of the current model would be to track the cold gas
content of each galaxy.  This would allow star formation to be limited to using
only that gas which is currently available and not already locked up in stars,
thus providing more realistic instantaneous star formation rates for individual
galaxies.  Merger-driven starbursts could additionally be implemented as
consuming some fraction of any available cold gas in the two progenitor
galaxies.  More advanced versions of this class of model have already been shown
to be successful in reproducing the results of full hydrodynamic simulations
\citep{Neistein2012}, and have been explored in other works using statistically
generated mass accretion histories \citep[e.g.][]{Bouche2010}.  However, it
is important to recognize that our aim with the formation history model is to
provide a simple, physically motivated, ``toy'' model.  By adding the ability to
track various reservoirs of material, or other similar complexities, we would
arrive at what is essentially a simplified semi-analytic galaxy formation model,
which is not the goal of this work.

\section{Conclusions}
\label{sec:conclusions}

In this work we introduce a simple model for self-consistently connecting
the growth of galaxies to the formation history of their host dark matter
haloes.  This is achieved by directly tying the time averaged change in
mass of a halo to the star formation rate of its galaxy via two simple
functions: the ``baryonic growth function'' and the ``physics function''
(Eqns.~\ref{eqn:bgf},\ref{eqn:physicsfunc_mvir}).  We utilize $N$-body dark matter
merger trees to provide self-consistent growth histories of individual haloes
that naturally includes scatter due to varying formation histories.  This allows
us to produce full star formation histories for individual objects, and thus
provide predictions for secondary properties such as galaxy colour.

While closely related to other models in terms of its basic methodology
\citep{Bouche2010,Cattaneo2011}, our model has a number of important
generalizations which enhance its utility.  In particular, we implement a
single, unified physics function which encapsulates the effects of all of the
intertwined baryonic processes associated with galactic star formation and
condenses them down into a simple mapping between star formation efficiency and
dark matter halo properties.  The qualitative form of this function is motivated
by our general knowledge of galaxy evolution, however, in this work we make no
attempt to directly tie it to individual physical processes or their particular
scalings with halo properties.

As well as introducing this new model, we demonstrate its ability to replicate
important observed relations such as the galactic stellar mass function, and
also illustrate some examples of its potential for investigating different
theories of galaxy formation and evolution.  Our main results can be summarized
as follows.
\begin{enumerate}
\item
  Motivated by the observed suppression of star formation efficiency in both the
  most massive and least massive dark matter haloes we begin by parametrizing the
  physics function as a simple, non-evolving, log-normal distribution with a
  single independent variable of either halo virial mass, $M_{\rm vir}$, or
  maximum circular velocity, $V_{\rm max}$ (Fig.~\ref{fig:cartoon}).
\item
  With just three free parameters controlling the position, normalization
  and dispersion of the peak star formation efficiency, we show that the
  formation history model can successfully reproduce the observed red and blue
  stellar mass functions at redshift zero.  Assuming a suitable choice of the
  parameters, this result is independent of the use of $M_{\rm vir}$ or $V_{\rm
  max}$ as the dependant variable of the physics function (Fig.~\ref{fig:z0_smf}).
\item
  For the purposes of replicating the stellar mass functions across a wide range
  of redshifts, we find our static model to be inadequate. This is due to its
  inability to produce the correct evolution of the stellar--halo mass relation
  with time (Figs~\ref{fig:3panel_smf}~and~\ref{fig:no_evo_SHMrelation}).
\item
  We therefore investigate the use of redshift as a second dependant
  variable to the physics function in order to control the position and
  normalization of the peak star formation efficiency with time. Using this
  simple adaptation alone, the formation history model is able to better
  reproduce the observed high-redshift stellar mass functions out to $z{=}3.5$
  (Figs~\ref{fig:evo_SHMrelation}~and~\ref{fig:evo_3panel_smf}) whilst still
  maintaining a good reproduction of the \z0{} colour-split \SMF{}.
\item
  By statistically calibrating the free model parameters using MCMC
  techniques throughout this work, we are able to use the marginalized posterior
  likelihood distributions to demonstrate the well behaved and transparent
  nature of our simple model (Fig.~\ref{fig:z0_smf_probs}).
\end{enumerate}

In order to demonstrate its construction and utility we have presented one of
the simplest forms of the formation history model.  However, a fundamental
strength of its construction is that it can be easily extended to arbitrary
levels of complexity in order to investigate a whole host of physical processes
associated with galaxy formation and evolution, some general examples of which
we have outlined in \S\ref{sec:discussion}.  In future work we will investigate
the predictions made when using alternative forms of the baryonic growth and
physics functions.  We will also extend the model to investigate the birth of
super-massive black holes and the evolution of the quasar luminosity function.

\section*{Acknowledgements}

Both SJM and GBP are supported by the ARC Laureate Fellowship of S. Wyithe.
SJM also acknowledges the support of a Swinburne University SUPRA postgraduate
scholarship.  DJC acknowledges receipt of a QEII Fellowship awarded by the
Australian government.

The authors would like to thank A. Knebe for useful discussions, as well
as the referee, E. Neistein, for numerous useful comments which have
helped to improve the content of this work.  The Millennium Simulation
used as input for the formation history model was carried out by the Virgo
Supercomputing Consortium at the Computing Centre of the Max-Planck Society
in Garching. Halo catalogues from the simulation are publicly available at
http://www.mpa-garching.mpg.de/millennium/

\bibliographystyle{mn2e}
\bibliography{master}

\begin{thebibliography}{66}
\expandafter\ifx\csname natexlab\endcsname\relax\def\natexlab#1{#1}\fi

\bibitem[{{Baldry} {et~al}\mbox{.}(2004){Baldry}, {Glazebrook}, {Brinkmann},
  {Ivezi{\'c}}, {Lupton}, {Nichol}, \& {Szalay}}]{Baldry2004}
{Baldry} I.~K., {Glazebrook} K., {Brinkmann} J., {Ivezi{\'c}} {\v Z}., {Lupton}
  R.~H., {Nichol} R.~C., {Szalay} A.~S., 2004, \apj, 600, 681

\bibitem[{{Baldry} {et~al}\mbox{.}(2008){Baldry}, {Glazebrook}, \&
  {Driver}}]{Baldry2008}
{Baldry} I.~K., {Glazebrook} K., {Driver} S.~P., 2008, \mnras, 388, 945

\bibitem[{{Behroozi} {et~al}\mbox{.}(2013{\natexlab{a}}){Behroozi}, {Wechsler},
  \& {Conroy}}]{Behroozi2013}
{Behroozi} P.~S., {Wechsler} R.~H., {Conroy} C., 2013{\natexlab{a}}, \apjl,
  762, L31

\bibitem[{{Behroozi} {et~al}\mbox{.}(2013{\natexlab{b}}){Behroozi}, {Wechsler},
  \& {Conroy}}]{Behroozi2013b}
{Behroozi} P.~S., {Wechsler} R.~H., {Conroy} C., 2013{\natexlab{b}}, \apj, 770,
  57

\bibitem[{{Bell} {et~al}\mbox{.}(2003){Bell}, {McIntosh}, {Katz}, \&
  {Weinberg}}]{Bell2003}
{Bell} E.~F., {McIntosh} D.~H., {Katz} N., {Weinberg} M.~D., 2003, \apjs, 149,
  289

\bibitem[{{Benson} {et~al}\mbox{.}(2002){Benson}, {Lacey}, {Baugh}, {Cole}, \&
  {Frenk}}]{Benson2002}
{Benson} A.~J., {Lacey} C.~G., {Baugh} C.~M., {Cole} S., {Frenk} C.~S., 2002,
  \mnras, 333, 156

\bibitem[{{B{\'e}thermin} {et~al}\mbox{.}(2012){B{\'e}thermin}, {Dor{\'e}}, \&
  {Lagache}}]{Bethermin2012}
{B{\'e}thermin} M., {Dor{\'e}} O., {Lagache} G., 2012, \aap, 537, L5

\bibitem[{{Binney} \& {Tremaine}(2008)}]{Binney2008}
{Binney} J., {Tremaine} S., 2008, {Galactic Dynamics: Second Edition}.
  Princeton University Press

\bibitem[{{Birnboim} \& {Dekel}(2003)}]{Birnboim2003}
{Birnboim} Y., {Dekel} A., 2003, \mnras, 345, 349

\bibitem[{{Bouch{\'e}} {et~al}\mbox{.}(2010){Bouch{\'e}}, {Dekel}, {Genzel},
  {Genel}, {Cresci}, {F{\"o}rster Schreiber}, {Shapiro}, {Davies}, \&
  {Tacconi}}]{Bouche2010}
{Bouch{\'e}} N. {et~al.}, 2010, \apj, 718, 1001

\bibitem[{{Bower} {et~al}\mbox{.}(2006){Bower}, {Benson}, {Malbon}, {Helly},
  {Frenk}, {Baugh}, {Cole}, \& {Lacey}}]{Bower2006}
{Bower} R.~G., {Benson} A.~J., {Malbon} R., {Helly} J.~C., {Frenk} C.~S.,
  {Baugh} C.~M., {Cole} S., {Lacey} C.~G., 2006, \mnras, 370, 645

\bibitem[{{Boylan-Kolchin} {et~al}\mbox{.}(2009){Boylan-Kolchin}, {Springel},
  {White}, {Jenkins}, \& {Lemson}}]{Boylan-Kolchin2009}
{Boylan-Kolchin} M., {Springel} V., {White} S.~D.~M., {Jenkins} A., {Lemson}
  G., 2009, \mnras, 398, 1150

\bibitem[{{Brooks} {et~al}\mbox{.}(2009){Brooks}, {Governato}, {Quinn},
  {Brook}, \& {Wadsley}}]{Brooks2009}
{Brooks} A.~M., {Governato} F., {Quinn} T., {Brook} C.~B., {Wadsley} J., 2009,
  \apj, 694, 396

\bibitem[{{Bruzual} \& {Charlot}(2003)}]{Bruzual2003}
{Bruzual} G., {Charlot} S., 2003, \mnras, 344, 1000

\bibitem[{{Cattaneo} {et~al}\mbox{.}(2008){Cattaneo}, {Dekel}, {Faber}, \&
  {Guiderdoni}}]{Cattaneo2008}
{Cattaneo} A., {Dekel} A., {Faber} S.~M., {Guiderdoni} B., 2008, \mnras, 389,
  567

\bibitem[{{Cattaneo} {et~al}\mbox{.}(2011){Cattaneo}, {Mamon}, {Warnick}, \&
  {Knebe}}]{Cattaneo2011}
{Cattaneo} A., {Mamon} G.~A., {Warnick} K., {Knebe} A., 2011, \aap, 533, A5

\bibitem[{{Cole} {et~al}\mbox{.}(2000){Cole}, {Lacey}, {Baugh}, \&
  {Frenk}}]{Cole2000}
{Cole} S., {Lacey} C.~G., {Baugh} C.~M., {Frenk} C.~S., 2000, \mnras, 319, 168

\bibitem[{{Cole} {et~al}\mbox{.}(2005){Cole}, {Percival}, {Peacock}, {Norberg},
  {Baugh}, {Frenk}, {Baldry}, {Bland-Hawthorn}, {Bridges}, {Cannon}, {Colless},
  {Collins}, {Couch}, {Cross}, {Dalton}, {Eke}, {De Propris}, {Driver},
  {Efstathiou}, {Ellis}, {Glazebrook}, {Jackson}, {Jenkins}, {Lahav}, {Lewis},
  {Lumsden}, {Maddox}, {Madgwick}, {Peterson}, {Sutherland}, \&
  {Taylor}}]{Cole2005}
{Cole} S. {et~al.}, 2005, \mnras, 362, 505

\bibitem[{{Conroy} \& {Wechsler}(2009)}]{Conroy2009}
{Conroy} C., {Wechsler} R.~H., 2009, \apj, 696, 620

\bibitem[{{Conroy} {et~al}\mbox{.}(2006){Conroy}, {Wechsler}, \&
  {Kravtsov}}]{Conroy2006}
{Conroy} C., {Wechsler} R.~H., {Kravtsov} A.~V., 2006, \apj, 647, 201

\bibitem[{{Conroy} {et~al}\mbox{.}(2007){Conroy}, {Wechsler}, \&
  {Kravtsov}}]{Conroy2007}
{Conroy} C., {Wechsler} R.~H., {Kravtsov} A.~V., 2007, \apj, 668, 826

\bibitem[{{Cowie} {et~al}\mbox{.}(1996){Cowie}, {Songaila}, {Hu}, \&
  {Cohen}}]{Cowie1996}
{Cowie} L.~L., {Songaila} A., {Hu} E.~M., {Cohen} J.~G., 1996, \aj, 112, 839

\bibitem[{{Croton} {et~al}\mbox{.}(2006){Croton}, {Springel}, {White}, {De
  Lucia}, {Frenk}, {Gao}, {Jenkins}, {Kauffmann}, {Navarro}, \&
  {Yoshida}}]{Croton2006}
{Croton} D.~J. {et~al.}, 2006, \mnras, 365, 11

\bibitem[{{Dekel} {et~al}\mbox{.}(2009){Dekel}, {Birnboim}, {Engel},
  {Freundlich}, {Goerdt}, {Mumcuoglu}, {Neistein}, {Pichon}, {Teyssier}, \&
  {Zinger}}]{Dekel2009b}
{Dekel} A. {et~al.}, 2009, \nat, 457, 451

\bibitem[{{Dekel} {et~al}\mbox{.}(2013){Dekel}, {Zolotov}, {Tweed}, {Cacciato},
  {Ceverino}, \& {Primack}}]{Dekel2013}
{Dekel} A., {Zolotov} A., {Tweed} D., {Cacciato} M., {Ceverino} D., {Primack}
  J.~R., 2013, ArXiv:1303.3009

\bibitem[{{Fontanot} {et~al}\mbox{.}(2009){Fontanot}, {De Lucia}, {Monaco},
  {Somerville}, \& {Santini}}]{Fontanot2009}
{Fontanot} F., {De Lucia} G., {Monaco} P., {Somerville} R.~S., {Santini} P.,
  2009, \mnras, 397, 1776

\bibitem[{Gelman \& Rubin(1992)}]{Gelman1992}
Gelman A., Rubin D.~B., 1992, Statistical Science, 7

\bibitem[{{Guo} {et~al}\mbox{.}(2011){Guo}, {White}, {Boylan-Kolchin}, {De
  Lucia}, {Kauffmann}, {Lemson}, {Li}, {Springel}, \& {Weinmann}}]{Guo2011}
{Guo} Q. {et~al.}, 2011, \mnras, 413, 101

\bibitem[{{Henriques} {et~al}\mbox{.}(2013){Henriques}, {White}, {Thomas},
  {Angulo}, {Guo}, {Lemson}, \& {Springel}}]{Henriques2013}
{Henriques} B.~M.~B., {White} S.~D.~M., {Thomas} P.~A., {Angulo} R.~E., {Guo}
  Q., {Lemson} G., {Springel} V., 2013, \mnras, 431, 3373

\bibitem[{{Jarrett} {et~al}\mbox{.}(2000){Jarrett}, {Chester}, {Cutri},
  {Schneider}, {Skrutskie}, \& {Huchra}}]{Jarrett2000}
{Jarrett} T.~H., {Chester} T., {Cutri} R., {Schneider} S., {Skrutskie} M.,
  {Huchra} J.~P., 2000, \aj, 119, 2498

\bibitem[{{Kauffmann} {et~al}\mbox{.}(1999){Kauffmann}, {Colberg}, {Diaferio},
  \& {White}}]{Kauffmann1999}
{Kauffmann} G., {Colberg} J.~M., {Diaferio} A., {White} S.~D.~M., 1999, \mnras,
  303, 188

\bibitem[{{Kennicutt}(1998)}]{Kennicutt1998}
{Kennicutt}, Jr. R.~C., 1998, \apj, 498, 541

\bibitem[{{Kere{\v s}} {et~al}\mbox{.}(2005){Kere{\v s}}, {Katz}, {Weinberg},
  \& {Dav{\'e}}}]{Keres2005}
{Kere{\v s}} D., {Katz} N., {Weinberg} D.~H., {Dav{\'e}} R., 2005, \mnras, 363,
  2

\bibitem[{{Krumholz} \& {Dekel}(2012)}]{Krumholz2012}
{Krumholz} M.~R., {Dekel} A., 2012, \apj, 753, 16

\bibitem[{{Li} \& {White}(2009)}]{Li2009}
{Li} C., {White} S.~D.~M., 2009, \mnras, 398, 2177

\bibitem[{{Lu} {et~al}\mbox{.}(2012){Lu}, {Mo}, {Katz}, \& {Weinberg}}]{Lu2012}
{Lu} Y., {Mo} H.~J., {Katz} N., {Weinberg} M.~D., 2012, \mnras, 421, 1779

\bibitem[{{Monaco} {et~al}\mbox{.}(2006){Monaco}, {Murante}, {Borgani}, \&
  {Fontanot}}]{Monaco2006}
{Monaco} P., {Murante} G., {Borgani} S., {Fontanot} F., 2006, \apjl, 652, L89

\bibitem[{{More} {et~al}\mbox{.}(2009){More}, {van den Bosch}, {Cacciato},
  {Mo}, {Yang}, \& {Li}}]{More2009}
{More} S., {van den Bosch} F.~C., {Cacciato} M., {Mo} H.~J., {Yang} X., {Li}
  R., 2009, \mnras, 392, 801

\bibitem[{{Moster} {et~al}\mbox{.}(2013){Moster}, {Naab}, \&
  {White}}]{Moster2013}
{Moster} B.~P., {Naab} T., {White} S.~D.~M., 2013, \mnras, 428, 3121

\bibitem[{{Mutch} {et~al}\mbox{.}(2011){Mutch}, {Croton}, \&
  {Poole}}]{Mutch2011}
{Mutch} S.~J., {Croton} D.~J., {Poole} G.~B., 2011, \apj, 736, 84

\bibitem[{{Mutch} {et~al}\mbox{.}(2013){Mutch}, {Poole}, \&
  {Croton}}]{Mutch2013}
{Mutch} S.~J., {Poole} G.~B., {Croton} D.~J., 2013, \mnras, 428, 2001

\bibitem[{{Neistein} {et~al}\mbox{.}(2012){Neistein}, {Khochfar}, {Dalla
  Vecchia}, \& {Schaye}}]{Neistein2012}
{Neistein} E., {Khochfar} S., {Dalla Vecchia} C., {Schaye} J., 2012, \mnras,
  421, 3579

\bibitem[{{Neistein} \& {Weinmann}(2010)}]{Neistein2010}
{Neistein} E., {Weinmann} S.~M., 2010, \mnras, 405, 2717

\bibitem[{{Papastergis} {et~al}\mbox{.}(2012){Papastergis}, {Cattaneo},
  {Huang}, {Giovanelli}, \& {Haynes}}]{Papastergis2012}
{Papastergis} E., {Cattaneo} A., {Huang} S., {Giovanelli} R., {Haynes} M.~P.,
  2012, \apj, 759, 138

\bibitem[{{Pe{\~n}arrubia} {et~al}\mbox{.}(2010){Pe{\~n}arrubia}, {Benson},
  {Walker}, {Gilmore}, {McConnachie}, \& {Mayer}}]{Penarrubia2010}
{Pe{\~n}arrubia} J., {Benson} A.~J., {Walker} M.~G., {Gilmore} G.,
  {McConnachie} A.~W., {Mayer} L., 2010, \mnras, 406, 1290

\bibitem[{Peacock \& Smith(2000)}]{Peacock2000}
Peacock J.~A., Smith R.~E., 2000, \mnras, 318, 1144

\bibitem[{{P{\'e}rez-Gonz{\'a}lez}
  {et~al}\mbox{.}(2008){P{\'e}rez-Gonz{\'a}lez}, {Rieke}, {Villar}, {Barro},
  {Blaylock}, {Egami}, {Gallego}, {Gil de Paz}, {Pascual}, {Zamorano}, \&
  {Donley}}]{Perez-Gonzalez2008}
{P{\'e}rez-Gonz{\'a}lez} P.~G. {et~al.}, 2008, \apj, 675, 234

\bibitem[{{Reddick} {et~al}\mbox{.}(2013){Reddick}, {Wechsler}, {Tinker}, \&
  {Behroozi}}]{Reddick2013}
{Reddick} R.~M., {Wechsler} R.~H., {Tinker} J.~L., {Behroozi} P.~S., 2013,
  \apj, 771, 30

\bibitem[{{Salpeter}(1955)}]{Salpeter1955}
{Salpeter} E.~E., 1955, \apj, 121, 161

\bibitem[{{Santini} {et~al}\mbox{.}(2012){Santini}, {Fontana}, {Grazian},
  {Salimbeni}, {Fontanot}, {Paris}, {Boutsia}, {Castellano}, {Fiore},
  {Gallozzi}, {Giallongo}, {Koekemoer}, {Menci}, {Pentericci}, \&
  {Somerville}}]{Santini2012}
{Santini} P. {et~al.}, 2012, \aap, 538, A33

\bibitem[{{Spergel} {et~al}\mbox{.}(2003){Spergel}, {Verde}, {Peiris},
  {Komatsu}, {Nolta}, {Bennett}, {Halpern}, {Hinshaw}, {Jarosik}, {Kogut},
  {Limon}, {Meyer}, {Page}, {Tucker}, {Weiland}, {Wollack}, \&
  {Wright}}]{Spergel2003}
{Spergel} D.~N. {et~al.}, 2003, \apjs, 148, 175

\bibitem[{{Springel} {et~al}\mbox{.}(2005){Springel}, {White}, {Jenkins},
  {Frenk}, {Yoshida}, {Gao}, {Navarro}, {Thacker}, {Croton}, {Helly},
  {Peacock}, {Cole}, {Thomas}, {Couchman}, {Evrard}, {Colberg}, \&
  {Pearce}}]{Springel2005}
{Springel} V. {et~al.}, 2005, \nat, 435, 629

\bibitem[{{Springel} {et~al}\mbox{.}(2001){Springel}, {White}, {Tormen}, \&
  {Kauffmann}}]{Springel2001}
{Springel} V., {White} S.~D.~M., {Tormen} G., {Kauffmann} G., 2001, \mnras,
  328, 726

\bibitem[{{Stoughton} {et~al}\mbox{.}(2002){Stoughton}, {Lupton}, {Bernardi},
  {Blanton}, {Burles}, {Castander}, {Connolly}, {Eisenstein}, {Frieman},
  {Hennessy}, {Hindsley}, {Ivezi{\'c}}, {Kent}, {Kunszt}, {Lee}, {Meiksin},
  {Munn}, {Newberg}, {Nichol}, {Nicinski}, {Pier}, {Richards}, {Richmond},
  {Schlegel}, {Smith}, {Strauss}, {SubbaRao}, {Szalay}, {Thakar}, {Tucker},
  {Vanden Berk}, {Yanny}, {Adelman}, {Anderson}, {Anderson}, {Annis},
  {Bahcall}, {Bakken}, {Bartelmann}, {Bastian}, {Bauer}, {Berman},
  {B{\"o}hringer}, {Boroski}, {Bracker}, {Briegel}, {Briggs}, {Brinkmann},
  {Brunner}, {Carey}, {Carr}, {Chen}, {Christian}, {Colestock}, {Crocker},
  {Csabai}, {Czarapata}, {Dalcanton}, {Davidsen}, {Davis}, {Dehnen},
  {Dodelson}, {Doi}, {Dombeck}, {Donahue}, {Ellman}, {Elms}, {Evans}, {Eyer},
  {Fan}, {Federwitz}, {Friedman}, {Fukugita}, {Gal}, {Gillespie}, {Glazebrook},
  {Gray}, {Grebel}, {Greenawalt}, {Greene}, {Gunn}, {de Haas}, {Haiman},
  {Haldeman}, {Hall}, {Hamabe}, {Hansen}, {Harris}, {Harris}, {Harvanek},
  {Hawley}, {Hayes}, {Heckman}, {Helmi}, {Henden}, {Hogan}, {Hogg}, {Holmgren},
  {Holtzman}, {Huang}, {Hull}, {Ichikawa}, {Ichikawa}, {Johnston}, {Kauffmann},
  {Kim}, {Kimball}, {Kinney}, {Klaene}, {Kleinman}, {Klypin}, {Knapp},
  {Korienek}, {Krolik}, {Kron}, {Krzesi{\'n}ski}, {Lamb}, {Leger},
  {Limmongkol}, {Lindenmeyer}, {Long}, {Loomis}, {Loveday}, {MacKinnon},
  {Mannery}, {Mantsch}, {Margon}, {McGehee}, {McKay}, {McLean}, {Menou},
  {Merelli}, {Mo}, {Monet}, {Nakamura}, {Narayanan}, {Nash}, {Neilsen},
  {Newman}, {Nitta}, {Odenkirchen}, {Okada}, {Okamura}, {Ostriker}, {Owen},
  {Pauls}, {Peoples}, {Peterson}, {Petravick}, {Pope}, {Pordes}, {Postman},
  {Prosapio}, {Quinn}, {Rechenmacher}, {Rivetta}, {Rix}, {Rockosi}, {Rosner},
  {Ruthmansdorfer}, {Sandford}, {Schneider}, {Scranton}, {Sekiguchi}, {Sergey},
  {Sheth}, {Shimasaku}, {Smee}, {Snedden}, {Stebbins}, {Stubbs}, {Szapudi},
  {Szkody}, {Szokoly}, {Tabachnik}, {Tsvetanov}, {Uomoto}, {Vogeley}, {Voges},
  {Waddell}, {Walterbos}, {Wang}, {Watanabe}, {Weinberg}, {White}, {White},
  {Wilhite}, {Wolfe}, {Yasuda}, {York}, {Zehavi}, \& {Zheng}}]{Stoughton2002}
{Stoughton} C. {et~al.}, 2002, \aj, 123, 485

\bibitem[{{Tacchella} {et~al}\mbox{.}(2013){Tacchella}, {Trenti}, \&
  {Carollo}}]{Tacchella2013}
{Tacchella} S., {Trenti} M., {Carollo} C.~M., 2013, \apjl, 768, L37

\bibitem[{{Wang} {et~al}\mbox{.}(2013){Wang}, {Farrah}, {Oliver}, {Amblard},
  {B{\'e}thermin}, {Bock}, {Conley}, {Cooray}, {Halpern}, {Heinis}, {Ibar},
  {Ilbert}, {Ivison}, {Marsden}, {Roseboom}, {Rowan-Robinson}, {Schulz},
  {Smith}, {Viero}, \& {Zemcov}}]{Wang2013}
{Wang} L. {et~al.}, 2013, \mnras, 431, 648

\bibitem[{{Weinmann} {et~al}\mbox{.}(2006){Weinmann}, {van den Bosch}, {Yang},
  {Mo}, {Croton}, \& {Moore}}]{Weinmann2006}
{Weinmann} S.~M., {van den Bosch} F.~C., {Yang} X., {Mo} H.~J., {Croton} D.~J.,
  {Moore} B., 2006, \mnras, 372, 1161

\bibitem[{{White} \& {Frenk}(1991)}]{White1991}
{White} S.~D.~M., {Frenk} C.~S., 1991, \apj, 379, 52

\bibitem[{{White} \& {Rees}(1978)}]{White1978}
{White} S.~D.~M., {Rees} M.~J., 1978, \mnras, 183, 341

\bibitem[{{Wisnioski} {et~al}\mbox{.}(2011){Wisnioski}, {Glazebrook}, {Blake},
  {Wyder}, {Martin}, {Poole}, {Sharp}, {Couch}, {Kacprzak}, {Brough},
  {Colless}, {Contreras}, {Croom}, {Croton}, {Davis}, {Drinkwater}, {Forster},
  {Gilbank}, {Gladders}, {Jelliffe}, {Jurek}, {Li}, {Madore}, {Pimbblet},
  {Pracy}, {Woods}, \& {Yee}}]{Wisnioski2011}
{Wisnioski} E. {et~al.}, 2011, \mnras, 417, 2601

\bibitem[{{Wyithe} \& {Loeb}(2007)}]{Wyithe2007}
{Wyithe} J.~S.~B., {Loeb} A., 2007, \mnras, 382, 921

\bibitem[{{Xue} {et~al}\mbox{.}(2008){Xue}, {Rix}, {Zhao}, {Re Fiorentin},
  {Naab}, {Steinmetz}, {van den Bosch}, {Beers}, {Lee}, {Bell}, {Rockosi},
  {Yanny}, {Newberg}, {Wilhelm}, {Kang}, {Smith}, \& {Schneider}}]{Xue2008}
{Xue} X.~X. {et~al.}, 2008, \apj, 684, 1143

\bibitem[{{Yang} {et~al}\mbox{.}(2009){Yang}, {Mo}, \& {van den
  Bosch}}]{Yang2009}
{Yang} X., {Mo} H.~J., {van den Bosch} F.~C., 2009, \apj, 693, 830

\bibitem[{{Yang} {et~al}\mbox{.}(2012){Yang}, {Mo}, {van den Bosch}, {Zhang},
  \& {Han}}]{Yang2012}
{Yang} X., {Mo} H.~J., {van den Bosch} F.~C., {Zhang} Y., {Han} J., 2012, \apj,
  752, 41

\bibitem[{{Zheng} {et~al}\mbox{.}(2005){Zheng}, {Berlind}, {Weinberg},
  {Benson}, {Baugh}, {Cole}, {Dav{\'e}}, {Frenk}, {Katz}, \&
  {Lacey}}]{Zheng2005}
{Zheng} Z. {et~al.}, 2005, \apj, 633, 791

\bibitem[{{Zheng} {et~al}\mbox{.}(2007){Zheng}, {Coil}, \&
  {Zehavi}}]{Zheng2007}
{Zheng} Z., {Coil} A.~L., {Zehavi} I., 2007, \apj, 667, 760

\end{thebibliography}

\label{lastpage}

\end{document}